\documentclass[12pt]{article}

\usepackage[T1]{fontenc}

\usepackage[thinspace,squaren,pstricks,italian,derivedinbase,derived]{SIunits}
\addunit{\bohr}{Bh}
\addunit{\rydberg}{Ry}
\addunit{\chwilka}{ch}
\addunit{\kubo}{Kb}

\usepackage[utf8]{inputenc}
\usepackage{amsfonts}
\usepackage{amsmath}
\usepackage{amssymb}
\usepackage{upgreek}  
\usepackage{mathrsfs}
\usepackage{booktabs} 
\usepackage{a4wide}
\usepackage{braket}
\usepackage{csquotes}

\newcommand{\ret}
{
^{\mathcal{R}}
}
\newcommand{\adv}
{
^{\mathcal{A}}
}

\newcommand{\tpr}
{
^{\mathcal{T}}
}
\newcommand{\crrl}
{
{\mathfrak{C}}
}

\newcommand{\pder}[2]
{
{\frac{\partial #1}{\partial #2}}
}

\renewcommand{\vec}[1]
{
{\mathbf #1}
}
\newcommand{\abs}[1]
{
{\left| #1 \right|} 
}

\newcommand{\kB}
{
{k_{\textrm{B}}}
}

\newcommand{\Bm}
{
{\mu_{\textrm{B}}}
}

\newcommand{\nno}
{
{\nonumber}
}
\newcommand{\eps}
{
{\epsilon}
}

\newcommand{\onemat}
{
{\sf I}     
}

\newcommand{\Heav}
{
{\theta}
}
\renewcommand{\Re}[1]
{
{\mathfrak{Re}\sqbr{#1}}
}

\newcommand{\ii}
{
{{\mathfrak{i}}} 
}
\newcommand{\ee}[1]
{
{\mathfrak{e}^{#1}}   
}

\newcommand{\av}[1]
{
{\left\langle #1 \right\rangle} 
}
\newcommand{\sqbr}[1]
{
{\left[ #1 \right]}
}
\newcommand{\cbr}[1]
{
{\left\lbrace #1 \right\rbrace}
}
\newcommand{\fbr}[1]
{
{\left( #1 \right)} 
}
\newcommand{\Complex}
{
{\mathbb{C}} 
}
\newcommand{\Reals}
{
{\mathbb{R}} 
}

\newcommand{\omb}[1]
{
{\omega_{#1}^{b}}
}

\newcommand{\zoneSymbol}
{
\Omega
}

\newcommand{\BZ}
{
\zoneSymbol_{\textrm{BZ}}
}

\newcommand{\suscSymb}
{
\chi
}

\newcommand{\comm}[3]
{
{\sqbr{#1,#2}_{#3}}
}

\newcommand{\Psigma}
{
{\pmb{\sigma}}
}

\newcommand{\fFD}
{
{f_{T}}
}

\newcommand{\latt}
{
\mathcal{L}
}

\newcommand{\zp}
{
{0^{+}}
}

\DeclareMathOperator{\tr}{tr} 

\DeclareMathOperator{\T}{T}
\newcommand{\tOrd}[2]
{
{\T_{#2}\sqbr{#1}}
}
\newcommand{\Ylm}
{
{\mathcal{Y}}
}

\newcommand{\xc}
{
_{\mathrm{xc}}
}
\newcommand{\KS}
{
_{\mathrm{KS}}
}
\newcommand{\KSs}
{
^{\mathrm{KS}}
}

\newcommand{\GSs}
{
^{\mathrm{GS}}
}

\newcommand{\el}
{%
\textit{et~al}.\
}

\newcommand{\vC}{v_{\textrm{C}}}

\usepackage{graphicx}
\usepackage[lf,p]{libertine}
\usepackage{inconsolata}

\usepackage[%
backend=biber,%
bibstyle=numeric,%
sortcites=true,%
maxcitenames=2,%
style=ACM-Reference-Format,%
citestyle=numeric-comp,%
sorting=none,%
defernumbers=true]{biblatex}

\begin{document}

\title{Dynamical magnetic susceptibility of non-collinear magnets: \\ 
a novel KKR-based \textit{ab initio} scheme and its application
}

\author{
David Eilmsteiner\\
\small{\textit{Institute for Theoretical Physics, Johannes Kepler University Linz, Altenberger  Stra{\ss}e 69, 4040 Linz}}\\
\small{\textit{Hamburg University of Applied Sciences, Berliner Tor 7, 20099 Hamburg, Germany}}\\
Arthur Ernst \\
\small{\textit{Institute for Theoretical Physics, Johannes Kepler University Linz, Altenberger  Stra{\ss}e 69, 4040 Linz}}\\
\small{\textit{Max Planck Institute of Microstructure Physics, Weinberg 2, 06120, Halle (Saale), Germany}}\\
\small{\textit{Donostia International Physics Center (DIPC), 20018 Donostia-San Sebasti\'{a}n, Spain}}\\
Pawe\l{} A. Buczek\footnote{Corresponding author: \texttt{pawel.buczek@haw-hamburg.de}}\\
\small{\textit{Hamburg University of Applied Sciences, Berliner Tor 7, 20099 Hamburg, Germany}}
} 

\date{\today}

\maketitle

\begin{abstract}
A novel implementation of the linear response time-dependent density functional theory addressing spin excitations in non-collinear magnets based on the Korringa-Kohn-Rostoker Green's function method is presented.
Following the exposition of the formalism based on the adiabatic local spin density approximation to the exchange-correlation kernel generalized to the noncollinear case, the computational scheme is discussed in detail.
The formation of the Goldstone modes in non-collinear susceptibility calculations is elaborated on formally and from the numerical convergence point of view.
The scheme is deployed to study the dispersion and Landau damping of magnons in the altermagnetic non-collinear kagome antiferromagnet Mn$_{3}$Ir.
The non-monotonous dependence of the damping on the magnon frequency makes the large momentum excitations attractive in the terahertz spintronics.
To this end, we analyze the real-time and real-space dynamics of the magnetic modes, including their strongly chirality-dependent attenuation.

\end{abstract}

\newpage
\tableofcontents
\newpage

\section{Introduction}

The \textit{ab initio} linear response time-dependent density functional theory (LRTDDFT) of spin excitations has become an indispensable tool in the computational design of novel functional materials 
\cite{%
Gross1985,%
Savrasov1998,%
Staunton1999,%
Buczek2009,%
Lounis2011,%
Skovhus2021,%
Skovhus2022,%
Durhuus2023,%
Binci2025}.
It offers a comprehensive description not only of spin-waves (magnons), which can be also studied by means of an empirical mapping onto the celebrated Heisenberg model \cite{LeBlanc2021},
but also of Stoner excitations and of the associated Landau damping,
inherent to the itinerant magnetism \cite{Moriya1985}.
No doubt, the dire need of the moment is the extension of the LRTDDFT to non-collinear (NC) magnets, i.e.,
spin-polarized solids where the direction of magnetization is allowed to vary as a function of position 
\cite{Sandratskii1998}.
Prompted to close this relevant methodological gap, in this paper, we present the first computer implementation of the LRTDDFT for NC magnets.

We are motivated by the recent developments in the spintronics and magnonics where an enormous practical potential has been ascertained to the members of this material family 
\cite{%
Cheong2024,Rimmler2025}.
Moreover, non-collinearity often accompanies other exotic or strongly correlated phases of matter, e.g., superconductivity
\cite{%
zlotnikovAspectsTopologicalSuperconductivity2021,%
dahlbergSpinglassDynamicsExperiment2025,%
Bruening2025%
}
and it naturally appears in the description of finite temperature magnetism \cite{Gyorffy1985,Antropov1996}.
However, the true attractive force of these magnets stems from their beautiful symmetries 
\cite{Sandratskii1991}
rooted deeply in the modern field theories
\cite{%
Azaria1993,%
Pradenas2024,%
Pradenas2025%
}.

A pivotal role in the development of magnonic devices is played by the controllability of the spin-waves' damping \cite{Chumak2015,Joshi2016,Duine2018}.
To begin with, they can scatter considerably on crystal imperfections
\cite{%
Buczek2016,%
Paischer2021,%
Paischer2021a,%
Paischer2024%
}.
Moreover, the NC magnets cannot be fully grasped within the linear spin-wave theory 
and their magnons undergo intrinsic damping due to two-boson interactions even at absolute zero 
\cite{Chernyshev2009,Zhitomirsky2013,Park2020}
contrary to the case of collinear ferromagnets \cite{Paischer2024}.
Finally, in metals, the collective density modes can interact with the single electron-hole pairs 
which leads to their possibly pronounced Landau decay mentioned above.
The mechanism was put forth by Landau to explain the plasmon attenuation \cite{Landau1946}
and stems from hybridization of the collective charge and spin density modes
with single electron-hole excitations.
In the case of collinear magnons, these states involve particles of opposite spins (up holes and down electrons for strong ferromagnets \cite{Paischer2024a}).
This simple picture is lost when the transitions stem from the NC band structure and the concept of the Stoner excitation must be revised \cite{Eilmsteiner2026a}.
This is achieved natively in the NC LRTDDFT formalism.

We apply our scheme to study the collective spin excitations in an important member of the frustrated NC kagome antiferromagnet (KAFM) family, Mn$_{3}$Ir \cite{Park2018,Park2020}.
Here, we are motivated by the fact that the kagome lattice gives rise to many remarkable properties, including Berry curvature, massive Dirac fermions, quantum Hall effect, topological states, and unconventional superconductivity
\cite{%
Kida2011,%
Ye2019,%
Chen2021,%
Jiang2021,%
Zhao2021,%
Guo2022%
}.
From the magnetism point of view, a long overlooked property of the KAFM is their inherently spin-polarized band structure persisting despite the vanishing net magnetization \cite{gurungNearlyPerfectSpin2024b,huSpinHallEdelstein2025},
akin to the altermagnetism reported in the collinear magnets
\cite{%
Mazin2022,%
Smejkal2022,%
cheongAltermagnetismNoncollinearSpins2024,%
mazinAltermagnetismThenNow2024,%
Maznichenko2024a,%
Sandratskii2025%
}.
It turns out that this NC altermagnetism gives rives to a series of novel magnonic effects, including exotic magnon chiralities and a polarization-dependent Landau damping \cite{Eilmsteiner2026a}.
A related phenomenon was reported in collinear compensated ferrimagnets \cite{Odashima2013} but has not been investigated in the NC case whatsoever.

The paper is organized as follows.
Sections \ref{sec:Formalism} and \ref{sec:Implementation} expose the underlying formalism and our computer implementation based on the Korringa-Kohn-Rostoker Green's function method.
In section \ref{sec:Results}, we concentrate on the spin density response of an archetypical KAFM, Mn$_{3}$Ir.
We discuss the magnon dispersion and the chirality-selective Landau damping, and proceed to the real-time and real-space analysis of the spin-wave modes, outlining their prospects in the the terahertz spintronics.

\section{Formalism}
\label{sec:Formalism}

\subsection{Generalized susceptibility}
\label{subsec:generalizedSusceptibility}

The following exposition pertains to a general interacting electron system with a possibly NC ground state.
Unless stated otherwise, Rydberg atomic units are used throughout, with $\hbar = 1$, Bohr radius $a_{0} = 1$, and Rydberg energy $E_{\mathrm{R}} = 1$ which in turn implies the numerical values of the electron charge and mass to be respectively $e = \sqrt{2}$ and $m_{e} = \frac{1}{2}$.

Here, the description of properties of spin and charge excitations is facilitated by the evaluation of the retarded density-density response function \cite{Fetter1971}
\begin{align}
  \chi^{ij} \fbr{\vec{x},\vec{x}',t - t'} &= - \ii \Heav(t - t')
    \av{\comm%
      {\hat{\sigma}^{i}\fbr{\vec{x}t}}
      {\hat{\sigma}^{j}\fbr{\vec{x}'t'}}
    {}}.
\label{eq:GeneralDensityResponse}
\end{align}
In the linear regime, the quantity yields the
charge or magnetization density response $\delta n^{i}\fbr{\vec{x},t}$ emerging due to the dynamical perturbing fields $\delta V^{j}\fbr{\vec{x}',t'}$ (magnetic or scalar) coupling to these densities.
$\hat{\sigma}^{i}\fbr{\vec{x}t}$ are charge ($i = 0$) and magnetization density operators ($i = x, y, z$), $\comm{A}{B}{}\equiv AB - BA$ is the commutator, and $\av{\hat{o}}$ is the expectation value of the operator $\hat{o}$ for the system without the perturbation. Throughout all the paper, $\vec{x}$ refers to the global position vector, whereas $\vec{r}$ is the position within the primitive unit cell. 
It is assumed that the potentials of the unperturbed system bear no time dependence in the Schr\"odinger picture.
Note that the response function can be used to study any electron density excitations including magnons, excitons, and plasmons.

The systems studied in this work are assumed to feature translational invariance which makes the response functions diagonal in the Bloch basis
\begin{align}
\chi\fbr{\vec{r}, \vec{r}', \vec{q}} = \sum_{\vec{R} \in \latt} 
    \ee{ - \ii \vec{q} \cdot \vec{R}}
    \chi\fbr{\vec{r} + \vec{R}, \vec{r}'}.
\label{eq:LatticeFourierTransformation}
\end{align}
Here, $\vec{r}$ and $\vec{r}'$ belong to the primitive cell of the crystal and $\latt$ denotes the set of lattice vectors. (In the equation above, other arguments are suppressed for simplicity.)

In the frequency domain, the anti-Hermitian part of the susceptibility (which we refer to as a \textit{loss matrix})
\begin{align}
L^{ij}\fbr{\vec{r}, \vec{r}', \vec{q}, \omega} \equiv \frac{1}{2\ii}
    \fbr{
        \chi^{ij}\fbr{\vec{r} , \vec{r}', \vec{q}, \omega} - 
        \chi^{ji}\fbr{\vec{r}', \vec{r} , \vec{q}, \omega}^{\ast}
    }.
\label{eq:lossMatrix}
\end{align}
allows to analyse the spectra and the spatial forms of the excited density modes.
For a specific energy $\omega$, the eigenvectors $\xi_{\lambda}^{i}\fbr{\vec{r}, \vec{q}, \omega}$ of $L$ represent the \emph{shapes} of the natural modes of charge and magnetization density fluctuations at this frequency.
The magnitude of the associated eigenvalues, $a_{\lambda}\fbr{\vec{q}, \omega}$, gives the density of states of these modes, up to a factor $ - 1/\pi$.
In other words, large $a_{\lambda}$ signifies the presence of density modes in the system.
These modes are real physical objects of the following form
\begin{align}
\Re{\xi_{\lambda}^{i}\fbr{\vec{r}, \vec{q}, \omega} 
    \ee{ \ii \fbr{\vec{q} \cdot \vec{R} - \omega t + \phi}}},
\end{align}
where $\phi$ is an overall phase of the mode.
When the frequency dependence of $a_{\lambda}\fbr{\vec{q}, \omega}$ is narrow,
one often identifies such modes as \enquote{collective}.
Following the \emph{fluctuation dissipation theorem}, which we briefly outline in Appendix \ref{app:KaellenLehmann}, the loss matrix has several further useful physical interpretations.
On one hand, $ - \omega a_{\lambda}\fbr{\vec{q}, \omega}/2$ yields the power absorbed by the system from the driving external fields of the shape $\xi_{\lambda}^{i}\fbr{\vec{r}, \vec{q}, \omega}$ coupling to the densities.
On the other, it can be used to compute the expectation values of the time correlations $\crrl^{ij}\fbr{\vec{r} , \vec{r}', \vec{q}, \omega}$ between density operators in the unperturbed state, i.e., among others, the spontaneous fluctuations of the system density.

Furthermore, the magnetization density loss matrix evaluated in momentum space is proportional to the cross-section of the inelastic neutron scattering off spin-density excitation \cite{VanHove1954}.
Last but not least, the quantity is related to the (spin-polarized) electron energy loss spectroscopy ((SP)EELS) signal
\cite{%
Kirschner1985,%
Vollmer2003,%
Costa2006,%
Santos2018}.
In the latter case, however, contrary to the neutron scattering, the loss energy dependence of the cross-section involves intricate details of the Coulomb interaction between the electrons of the beam and the probe
\cite{%
Vignale1985,%
Plihal1998,%
Hong2000a,%
Paischer2024b}.

\subsection{Linear response time dependent density function theory formulation}
\label{subsec:lrtddftFormulation}

The evaluation of the dynamic susceptibilities of an interacting electron gas is as involved as the solution of the many-body problem itself \cite{Abrikosov1975}. Consequently, one must resort to approximations.
A successful \textit{ab initio} scheme can be derived within the LRTDDFT 
\cite{Gross1985,Qian2002}. It often involves two steps.
First, the so called \textit{Kohn-Sham susceptibility}, $\suscSymb\KS$ is constructed describing the response of the formally non-interacting Kohn-Sham (KS) system to an external perturbation.
Second, the self-consistent dynamical change in the exchange-correlation potential arising due to the densities induced by the external field.
This change is incorporated upon solving the so called \textit{susceptibility Dyson equation}.
There are different complementary approaches to implement this scheme of which some do not explicitly break it into these two distinct phases \cite{Savrasov1998,Binci2025}.
The magnetic response function can be also found within the many-body perturbation scheme \cite{Aryasetiawan1999,Karlsson2000,Mueller2016,Nabok2021} and using empirical Hubbard-type Hamiltonians \cite{Cooke1980,Gumbs1980,Tang1998,Muniz2002,Bonetti2022}.

Here, we discuss the form of the formalism underlying our implementation.
The KS system is described by a corresponding formally single-electron Green's function (GF)
\begin{align}
  G_{\alpha\beta}\fbr{\vec{x},\vec{x}',z} = \sum_{j}
    \frac{\phi_{j}\fbr{\vec{x}\alpha}\phi_{j}\fbr{\vec{x}'\beta}^{\ast}}
      {z - \eps_{j}}, \, z\in\Complex,
\label{eq:BareGreensFunction}
\end{align}
where $\phi_{j}\fbr{\vec{x}\alpha}$'s and $\eps_{j}$'s stand for KS eigenfunctions and eigenenergies, respectively, and $\alpha, \beta, \ldots$ denote spin degrees of freedom (up or down along a chosen quantization axis).
The charge and magnetization densities, necessary for the Kohn-Sham potential's construction, can be found by complex integration of $G$.

%
%

In the Korringa-Kohn-Rostoker method \cite{Mankovsky2011,Simon2019}, the Green's function is not constructed from this K\"all\'e{}n-Lehmann spectral representation but found directly upon solving the Dyson equation, written formally as
\begin{align}
  G\fbr{z} = G_{0}\fbr{z} + 
      G_{0}\fbr{z} V\fbr{z} G\fbr{z}.
\label{eq:KKRDysonEquation}
\end{align}
(For simplicity the arguments of the Green's functions as well as corresponding sums and integrals are suppressed.) The ground state potential $V$ includes static external fields as well as the ground state Hartree and exchange-correlation contributions.
In this case, it does not feature the dependence on the complex energy $z$ and is local.
This is the form used to describe the materials considered in this paper.

Additionally, in many cases, it is necessary to include terms originating from the spin-orbit coupling (SOC) \cite{Paischer2024}.
Furthermore, the frequency dependence arises when typical electronic self-energy terms are incorporated, as it is the case when the disorder is described in the coherent potential approximation (CPA) \cite{Buczek2016,Paischer2021,Paischer2021a} or many-body effects like magnon-electron interaction are included \cite{Paischer2023,Usachov2024,Paischer2024a}.
In the latter case, the self-energy $V$ can additionally become non-local.

With the KS GF at hand, one can apply the Cauchy theorem in order to obtain the retarded KS susceptibility analytically continued to a line in a finite distance $\gamma$ from the real axis (the so called \textit{nearly real axis})
\cite{Abrikosov1975,
Schmalian1996}
($\gamma > \gamma' = \zp$):
\begin{align}
  \chi^{ij}\KS&\fbr{\vec{x},\vec{x}',\omega + \ii\gamma}
  = - \frac{1}{2\pi\ii} \int_{-\infty}^{\infty} d\eps \times \nno\\
    (
       &\fFD\fbr{\eps + \ii\gamma'}
        S_{ij}\fbr{\vec{x},\vec{x}',\eps + \omega + \ii\gamma + \ii\gamma',\eps + \ii\gamma'}\nno\\
      -&\fFD\fbr{\eps - \ii\gamma'}
        S_{ij}\fbr{\vec{x},\vec{x}',\eps + \omega + \ii\gamma - \ii\gamma',\eps - \ii\gamma'}\nno\\
      +&\fFD\fbr{\eps + \ii\gamma'}
        S_{ij}\fbr{\vec{x},\vec{x}',\eps + \ii\gamma',\eps - \omega - \ii\gamma + \ii\gamma'}\nno\\
      -&\fFD\fbr{\eps - \ii\gamma'}
        S_{ij}\fbr{\vec{x},\vec{x}',\eps - \ii\gamma',\eps - \omega - \ii\gamma - \ii\gamma'}
    )
\label{eq:Schmalian}
\end{align}
where the $S$ is obtained as a trace of the product of two GFs and Pauli matrices $\sigma^{i}$.
\begin{align}
  S_{ij}\fbr{\vec{x},\vec{x}',z_{1},z_{2}} \equiv
  \sigma^{i}_{\alpha\beta}
    G_{\beta\gamma}
        \fbr{\vec{x} ,\vec{x}',z_{1}}
    \sigma^{j}_{\gamma\delta}
    G_{\delta\alpha}
        \fbr{\vec{x}',\vec{x} ,z_{2}}.
\label{eq:Sij}
\end{align}
$\fFD(\eps_{j})$ stands for the Fermi-Dirac distribution function. 
A small finite temperature is introduced to slightly smear the Fermi level allowing for better convergence of the Brillouin zone integrals.
In this work, this temperature is negligible on the electronic energy scales and the electron gas is regarded to be described at absolute zero. 
$\zp$ denotes the small positive energy limit.
Einstein summation convention is assumed for the spin indices. There is no difference raised and lowered density index $i$.
The advantage of this approach is the possibility of deforming the complex plane integration contours such that the evaluation of $S$ takes place far from the singularities of the KS Green's function allowing for its more rapid numerical convergence \cite{Buczek2011a}.
Furthermore, the evaluation of the susceptibility on the nearly real axis yields a more stable analytic continuation to the real frequencies and magnon line fitting schemes than the evaluation at Matsubara poles
\cite{%
Staunton1999,%
Staunton2000,%
Thakor2003}
although it is computationally more expensive.

The susceptibility Dyson equation allows to find the true interacting (enhanced) susceptibility of the many-electron system:
\begin{align}
  \chi^{ij} &\fbr{\vec{x},\vec{x}',\omega} = \chi^{ij}\KS \fbr{\vec{x},\vec{x}',\omega} + \\ \nno
    &\sum_{k,l=0}^{3}
    \iint d\vec{x}_{1}d\vec{x}_{2} \chi^{ik}\KS \fbr{\vec{x},\vec{x}_{1},\omega}
      \fbr{
        \frac{2 \delta_{k0}\delta_{l0}}{\abs{\vec{x}_{1} - \vec{x}_{2}}} +
        K\xc^{kl}\fbr{\vec{x}_{1},\vec{x}_{2},\omega}
      }
      \chi^{lj}\fbr{\vec{x}_{2},\vec{x}',\omega}
\label{eq:susceptibilityDysonEquation}.
\end{align}
The first term in the bracket above denotes the Hartree (Coulomb) interaction ($\vC\fbr{\vec{x}} \equiv 2/x$) while
$K\xc$ stands for the exchange-correlation kernel, defined as a functional derivative of the exchange-correlation potential evaluated at the ground state values of electronic and magnetic densities
\begin{align}
  K\xc^{ij}\sqbr{\av{\hat{\Psigma}\fbr{\vec{x}}}}\fbr{\vec{x},\vec{x}',t-t'} \equiv
    \frac{\delta v\xc^{i}\fbr{\vec{x},t}}{\delta n^{j}\fbr{\vec{x}'t'}}.
\end{align}

\subsection{Exchange-correlation kernel}
\label{subsec:xcKernel}

The exact form of the $K\xc$ is, in general, unknown, especially for NC systems. There is a steady progress in constructing NC spin density functionals
\cite{%
Sharma2007,%
Kurth2009,%
Eich2010,%
Eich2013,%
Eich2013a,%
Pittalis2017,%
Eich2018}, but their their inclusion in practical LRTDDFT schemes is still a work in progress.
In this work, we adopt the adiabatic local spin density approximation (ALSDA) yielding $K\xc$
in the following form
\begin{align}
  K\xc^{ij}\sqbr{\av{\hat{\Psigma}\fbr{\vec{x}}}}\fbr{\vec{x},\vec{x}',t - t'} \approx
    \frac{\delta v_{\textrm{LSDA}}^{i}\sqbr{\av{\hat{\Psigma}\fbr{\vec{x}}},\vec{x}}}{\delta n^{j}\fbr{\vec{x}}}
    \delta\fbr{\vec{x} - \vec{x}'} \delta\fbr{t - t'}.
\end{align}

For NC systems, this kernel is most easily written in a local coordinate system with its $z$-axis oriented along the magnetization direction
\cite{vonBarth1972,Katsnelson2004}.
In this case, its space-diagonal part reads
\begin{align}
K\xc^{\textrm{loc}} = 
    \begin{pmatrix}
        \pder{V\xc}{n}        & 0 & 0 &        \pder{V\xc}{m}         \\
        0 &  K^{\perp} & 0         & 0         \\
        0                     & 0                     & K^{\perp} & 0         \\ 
         - \Bm \pder{B\xc}{n}                     & 0                     & 0         &  - \Bm \pder{B\xc}{m} \\ 
\end{pmatrix}
\end{align}
The matrix above is indexed with $i,j=0,x,y,z$,
the spatial argument $\vec{x}$ of the kernel has been omitted for brevity in 
$K\xc^{\textrm{loc}}\fbr{\vec{x}}$ and in the partial derivatives on the right hand side.
The derivatives are taken at the ground state values of the densities.
$V\xc\fbr{\vec{x}}$ and $B\xc\fbr{\vec{x}}$ are the local magnitudes of the exchange-correlation scalar potential and magnetic field. 
We note the symmetry
\begin{align}
\pder{V\xc}{m} = - \Bm \pder{B\xc}{n}
\end{align}
and the fact that
\begin{align}
K^{\perp}\fbr{\vec{x}} &= - \Bm 
    \frac{
        B\GSs_{\gamma}\fbr{\vec{x}}%
    }{
        m\GSs_{\gamma}\fbr{\vec{x}}
    }.
\label{eq:perpKxc}
\end{align}
$K^{\perp}$ is found simply as the ratio of the magnitudes of KS magnetic field and magnetization density.
We recall that these two latter vectors are parallel everywhere in the LSDA.
Furthermore, it is immediately evident that $K^{\perp}$ is isotropic in the subspace of the
transverse magnetization direction.
This form has a clear physical interpretation.
Once a infinitesimal magnetization is induced in the direction perpendicular to $\vec{m}\GSs\fbr{\vec{x}}$
the local $\vec{B}\xc\fbr{\vec{x}}$ rotates locally without changing its magnitude.

$K\xc^{\textrm{loc}}$ is brought into the global coordinate system using a rotation matrix
\begin{align}
K\xc = \mathcal{T}  K\xc^{\textrm{loc}} \mathcal{T}^{\dagger}
\end{align}
given by
\begin{align}
\mathcal{T} = 
    \begin{pmatrix}
        1 & 0                        & 0             & 0                        \\
        0 & \cos\vartheta\cos\varphi & - \sin\varphi & \sin\vartheta\cos\varphi \\
        0 & \cos\vartheta\sin\varphi &   \cos\varphi & \sin\vartheta\sin\varphi \\ 
        0 &  - \sin\vartheta         & 0             & \cos\vartheta 
    \end{pmatrix}
\end{align}
with $\vartheta(\vec{x})$ and $\varphi(\vec{x})$ being the polar and azimuthal angles describing the local direction of the spin magnetization at point $\vec{x}$. For simplicity, this spatial argument is omitted in the matrix above.

\subsection{The emergence of the Goldstone modes in noncollinear magnets}
\label{subsec:emergenceOfTheGoldstoneModesInNonCollinearMagnets}

The appearance of Nambu-Goldstone bosons (NGBs) modes due to the spontaneous breaking of continuous symmetries and the resulting presence of the degenerate manifold of ground-states is a well studied phenomenon
\cite{%
Nambu1960,%
Goldstone1961,%
Goldstone1962}.
Magnetic orderings are its manifestations, too. In the case of collinear magnets, the original symmetry of the Hamiltonian, $\mathrm{SO}\fbr{3}$, spontaneously breaks down to $\mathrm{SO}\fbr{2}$
\cite{%
Anderson1990,%
Peierls1991,%
Tscheuschner1992}.
The broken symmetry generators are spin angular momentum components, $S_{i}$, the magnetic order parameter transform as a vector, and the manifold of vacua is the two-dimensional hypersphere.

However, the question of the actual number of NGBs and their dispersion in systems without Lorentz invariance featuring vacuum expectation values of field-theoretical charges is a more delicate one
\cite{%
Brauner2007,%
Watanabe2012}.
Interestingly, as discussed by Nambu \cite{Nambu2004}, even the familiar cases of ferro- and antiferromagnetic orderings yield here two remarkably different pictures.
In both cases there are two continuous broken symmetry generators, say $S_{x}$ and $S_{y}$, but in ferromagnets they become canonically conjugated yielding only one NGB mode with quadratic dispersion relation.
On the contrary, in antiferromagnets, the two corresponding NGBs remain independent and both feature linear dispersion.

The spontaneous symmetry breaking in the case of a triangular antiferromagnet (TAF) is even more complex.
The magnetic order parameter transforms as a three-dimensional body.
As carefully discussed by Pradenas \el \cite{Pradenas2025}, the full symmetry of the TAF Hamiltonian,
$\mathrm{SO}\fbr{3}_{L} \times \mathrm{SO}\fbr{2}_{R,3}$,
closely related to the one of a axially-symmetric rigid quantum top \cite{Azaria1993},
is spontaneously reduced to the residual symmetry $\mathrm{SO}\fbr{2}_{V,3}$.
Three NGB branches with linear dispersion emerge \cite{Chernyshev2009}.
The residual symmetry leads to the conservation of the isospin and degeneracy of two spin-wave velocities.
It is further broken down when long-range interactions between spins and the inter-plane magnetic coupling are taken into account, lifting the degeneracy of the two magnon branches \cite{LeBlanc2014,LeBlanc2021}.
 
While the emergence of the three Goldstone bosons in NC triangular ferromagnets can be be deduced from the general symmetry arguments outlined above, it is instructive to inquire how it manifests itself in the LRTDDFT calculations.
Here, the correct long-wavelength behaviour, i.e., the presence of a mode or modes with vanishing energy for $q = 0$, can be seen as a sum rule arising from the interplay between the exchange correlation kernel $K\xc$ and the static uniform magnetic susceptibility
\cite{%
Lounis2010,%
Lounis2011,%
Buczek2011a,%
Mueller2016}.
We name this interplay \textit{Goldstone sum rule} and it can be stated as follows.
The \enquote{Dyson denominator} of Eq.~\eqref{eq:susceptibilityDysonEquation} for $q = 0$ and $\omega = 0$
(Coulomb interaction, $\vC$, plays no role in the formation of the magnetic NGB with anisotropy neglected)
\begin{align}
\onemat - \suscSymb\KS K\xc
\end{align}
must feature zero eigenvalues.
Their multiplicity, $N_{r}$, is equal to the number of linearly independent rigid rotations of the magnetic order parameter which, with spin-orbit coupling neglected, cost no energy.
We refer to them as \textit{zero-energy magnetic deformations}.
They are the eigenvectors of the Dyson denominator with the eigenvalue 0.
In a NC magnets $N_{r} = 3$ while $N_{r} = 2$ holds for the collinear case, i.e., it is directly given by the number of broken symmetry generators.
However, in the light of the preceding discussion, we note that $N_{r}$ is not necessarily equal to the number of Goldstone modes of a system.

Let us now turn to our specific case of the ALSDA.
Apart from having no frequency dependence, this kernel, cf. Sec.~\ref{subsec:xcKernel}, is $q$-independent and finite at $q = 0$.
The latter property tends to be a serious drawback in the charge-charge response channel, as it precludes the correct TDDFT description of excitonic effects in semiconductors \cite{Onida2002,Turkowski2009}.
On the other hand, it is the correct behaviour in the long wave-length limit for magnetization density channels, yielding correctly the magnon Goldstone mode which would not arise otherwise.

Generally speaking, there are two reasons why the Goldstone sum rule is not fulfilled exactly in numerical calculations.

The first is of systematic character and arises often in the many-body perturbation methods \cite{Mueller2016,Nabok2021} where the ground state found self-consistently using an approach (typically LSDA to the ground state functional) different from the one for the kernel (corresponding to the screened Coulomb interaction $W$).
Once $W$ were used to self-consistently find the magnetic ground state the deficiency would disappear.
Let us briefly remark that the currently available many-body perturbation schemes for computing magnetic excitations are practically equivalent to the LRTDDFT in the ALSDA.
They take advantage of the LSDA for the electronic ground state, local and static approximation to the screened Coulomb interaction $W$, as well as the choice of Feynman diagrams yielding the random phase approximation-like equation for the dynamic susceptibility resembling formally the equation \eqref{eq:susceptibilityDysonEquation} of the LRTDDFT.

The second reason are numerical inaccuracies arising during the computations of $\chi$.
They can be broadly said to have two origins.
The first includes the convergence of $\chi\KS$ calculations, typically with respect to the Brillouin zone and energy integration as well as number of KS states used.
Note that the latter aspect is of minor importance for the KKR-based implementations, contrary to the methods using direct K\"all\'en-Lehmann spectral representation for $\chi\KS$ evaluation.
The second origin of numerical errors is the finite or imperfect basis for the spatial representation of $\chi\KS$ and $\chi$.
In any case, accurate fulfilment of the Goldstone sum rule is crucial to the faithful description of magnon in the long wave-length regime and we look in detail now how it can be achieved in the case of the NC ground state and ALSDA.

Let us assume that $m\GSs_{\alpha a}$ and $B\GSs_{\alpha a}$ are ground state Kohn-Sham magnetic field and the magnetization density, respectively.
Here, $\alpha$ encodes the vector direction and $a$ the spatial dependence of the quantities, i.e., the position, $a\equiv\fbr{\vec{x}}$.
The zero-energy magnetic deformation of the ground state can be written as
\begin{align}
m^{\beta}_{\alpha a} =
    \phi \sum_{\gamma}
    \epsilon_{\alpha\beta\gamma}
    m\GSs_{\gamma a}
\end{align}
where $\epsilon_{\alpha\beta\gamma} \equiv \fbr{\beta - \alpha}\fbr{\gamma - \alpha}\fbr{\gamma - \beta}/2$ is the Levi-Civita symbol and $\phi$ an infinitesimal rotation angle of the ground state magnetization around the direction $\beta$. ($\alpha,\beta,\gamma \in \Set{1,2,3} \equiv \Set{x,y,z}$). Note that a linear combination of vectors $m^{\beta}_{\alpha a}$ corresponding to different $\beta$'s is a zero-energy magnetic deformation corresponding to an infinitesimal rotation around a specific axis as well. $m^{\beta}_{\alpha a}$ can be named \enquote{transverse} magnetic deformation because its direction (in the magnetization space) is orthogonal to the one of $m\GSs_{\alpha a}$. However, in the NC magnets, it does not lie in a single plane as in the collinear case.

When the deforming transverse magnetic field 
\begin{align}
B^{\beta}_{\alpha a} =
    \phi \sum_{\gamma}
    \epsilon_{\alpha\beta\gamma}
    B\GSs_{\gamma a}
\end{align}
is applied, the effective field of the Kohn-Sham system will rotate around $\beta$ by the angle $\phi$, too. When the Kohn-Sham equations are solved with this new field the resulting change of magnetization will become $m^{\beta}_{\alpha a}$ such that
\begin{align}
m^{\beta}_{\alpha a} =  - \Bm
    \sum_{\gamma b} 
    \chi\KSs_{\alpha a \gamma b}
        \fbr{\vec{q} = \vec{0}, \omega = 0}   
    B^{\beta}_{\gamma b}.
\end{align}
$\chi\KSs_{\alpha a \gamma b}\fbr{\vec{q} = \vec{0}, \omega = 0}$ is a Hermitian matrix and thus features a complete set of eigenvectors.
While the deformations $m^{\alpha}$ and $B^{\alpha}$ are clearly not eigenvectors of $\chi\KSs$, they can be decomposed in the eigenbasis of $\chi\KSs$.

It appears that only a part (which we will refer to as \enquote{transverse subspace}) of the eigenbasis is necessary to decompose $m^{\alpha}$ and $B^{\alpha}$.
Furthermore, all eigenvalues of this subspace are $N_{r}$-times degenerate and only one vector from the linear space spanned by the $N_{r}$ corresponding eigenvectors enters the decomposition. The latter can be written as
\begin{align}
B^{\beta}_{\alpha a} = 
\sum_{\lambda}
    b^{\lambda}
    v^{\lambda\beta}_{\alpha a}
=
\sum_{\lambda}
    b^{\lambda}
    f^{\lambda}_{a}
    B^{\beta}_{\alpha a}.
\label{eq:bDecomp}
\end{align}
We observe that the respective eigenvectors of the transverse subspace have the same magnetization direction as the deforming transverse magnetic fields but different magnitudes locally in space given by the factors $f^{\lambda}_{a}$.
Note that 
$\sum_{\lambda}
    b^{\lambda}
    f^{\lambda}_{a} = 1$
at every point in space.
The decomposition of the magnetization reads
\begin{align}
m^{\beta}_{\alpha a} = 
\sum_{\lambda}
    w^{\lambda}
    b^{\lambda}
    f^{\lambda}_{a}
    B^{\beta}_{\alpha a}.
\label{eq:mDecomp}
\end{align}
Eqs.~\eqref{eq:bDecomp} and \eqref{eq:mDecomp} allow to write the transverse part of the ex-kernel as
\begin{align}
K^{\perp}_{a} = - \Bm
\frac{B^{\beta}_{\alpha a}}
     {m^{\beta}_{\alpha a}}
= - \Bm
\frac{\sum_{\lambda}
        b^{\lambda}
        v^{\lambda\beta}_{\alpha a}
    }
    {\sum_{\lambda}
        w^{\lambda}
        b^{\lambda}
        v^{\lambda\beta}_{\alpha a}
    }
= - \Bm
\frac{1}
    {\sum_{\lambda}
        w^{\lambda}
        b^{\lambda}
        f^{\lambda}_{a}
    }.
\end{align}
One notes that the expression above is equivalent to Eq.~\eqref{eq:perpKxc} as
$
\sum_{\lambda}
        w^{\lambda}
        b^{\lambda}
        f^{\lambda}_{a}
$        
is the inverse ratio of the magnitudes of KS magnetic field and magnetization density which in the ALSDA point always in the same direction.

The discussion above allows to construct $K^{\perp}$ fulfilling the Goldstone sum rule exactly.
Instead of computing $K^{\perp}$ from $m\GSs$ and $B\GSs$ using Eq.~\eqref{eq:perpKxc} one can find $b^{\lambda}$ from the the decomposition of one of the pairs $\fbr{m^{\alpha},B^{\alpha}}$ in the eigenvectors of the transverse subspace.

Here, we remark briefly that in our new scheme, with the improved spatial basis, providing the convergence with respect to the Brillouin zone and energy integrations is sufficient, this correction is hardly necessary as the Goldstone sum rule is fulfilled with accuracy of few meV for $q = 0$ and the relative error of the pertinent degeneracy of eigevalues is about $10^{-5}$.

Let us conclude this section with the remark that the degeneracy $N_{r} = 2$ is trivially given in the collinear case where the transverse part of the $q = 0$ and $\omega = 0$ susceptibility is block diagonal and isotropic
\begin{align}
\begin{pmatrix}
    \chi\KSs_{xaxb} & 0               \\
    0               & \chi\KSs_{yayb}
\end{pmatrix}, \quad \chi\KSs_{yayb} = \chi\KSs_{xaxb}
\end{align}
The corresponding transverse subspace is associated with the magnetization tilting around directions perpendicular to the ground state magnetization. (In our case, they lie in $x$-$y$ plane). In every spin channel independently, $v^{\lambda}_{a}$ are simply eigenvectors of the
$\chi\KSs_{xaxb}$ block.

\section{Implementation}
\label{sec:Implementation}

\subsection{Korringa-Kohn-Rostoker Green's function method}
\label{subsec:KKRGF}

The generalization of the Korringa-Kohn-Rostoker (KKR) Green's function (GF) method to NC ground states is based on the following approach \cite{Mankovsky2011,Simon2019}.
First, the single site scattering problem is solved yielding the corresponding regular and irregular solutions, $R_{\sigma L \sigma' L'}^m(r_m,z)$ and $H_{\sigma L \sigma' L'}^m(r_m,z)$  as well as the corresponding $t$-matrix.
Subsequently, the backscattering operator taking into account the multiple electron-scattering in the lattice is found as

\begin{align}
	\mathcal{G}_{\sigma L \sigma' L'}^{mn}(z)=g_{LL'}^{mn}(z)\delta_{\sigma\sigma'}+g_{LL_1}^{mk}(z)t_{\sigma L_1 \sigma_2 L_2}^{k}(z)\mathcal{G}_{\sigma_2 L_2 \sigma' L'}^{kn}(z)
	\label{eq:bsc}
\end{align}
where $g_{LL'}^{mn}(z)$ is the reference Green's function. This allows to write down the KKR Green's function as

\begin{align}
G_{\sigma\sigma'}(\vec{x},\vec{x}',z)=&\Bigl[R_{\sigma L \sigma_1 L1}^m(r_m,z)\mathcal{G}_{\sigma_1 L_1 \sigma_2 L_2}^{mn}(z)R_{\sigma_2 L_2 \sigma' L'}^n(r_n,z)\nno \\ +&\sqrt{z}\delta_{mn}R_{\sigma L \sigma_1 L_1}^m(r_m^<,z)H_{\sigma_1 L_1 \sigma' L'}^m(r_n^>,z)\Bigr]\Ylm_{L}(\hat{\vec{r}}_m)\Ylm_{L'}(\hat{\vec{r}}_n)
\label{eq:green}
\end{align}
Einstein sum convention is used in equations \eqref{eq:bsc} and \eqref{eq:green}.
The KKR approach yields an all-electron picture of the Kohn-Sham system with the core states and atomic potentials found self-consistently.
We assume the system to feature the discrete translational symmetry including the magnetization orientation, thus, 
we do not have to deploy the generalized Bloch theorem for incommensurate spirals in the spirit of Sandratskii \cite{Sandratskii1991}.

Our KKR scheme is capable of describing full scattering potentials, i.e., of arbitrary non-spherical shape, including varying the direction of magnetization within an atomic cell (intra-atomic NCity) \cite{Papanikolaou2002}.
To this end, the shape functions \cite{Zeller1987} are used to define the Wigner-Seitz polyhedra \cite{Stefanou1990} both when computing the magnetization and charge densities as well as when evaluating the overlap integrals in the susceptibility Dyson equation \eqref{eq:susceptibilityDysonEquation}.
However, the closed lattice structure of the systems under consideration allows to resort to the atomic sphere or full-charge density approximation.
Furthermore, we observe that the intra-atomic NCity in the systems under consideration in this work is practically negligible \cite{Sharma2007} and, thus, assume that the magnetization points along a certain common direction within the given atomic polyhedron $s$ defined by the angles $\vartheta_{s}$ and $\varphi_{s}$.


\subsection{Evaluation of susceptibility}
\label{subsec:evaluationOfSusceptibility}

We turn now to the details of the evaluation of the density response functions.
The complex plane integration scheme allowing to evaluate Eq. \eqref{eq:Schmalian} is exposed in detail in \cite{Buczek2011a}.
It does not differ from the collinear case.
The spatial $\fbr{\vec{x}, \vec{x'}}$ dependence of $\suscSymb$ is treated as follows.
The discrete periodicity of the atomic system, if any, is used to diagonalize the susceptibility in the Bloch basis and cast it into the form of $\suscSymb\fbr{\vec{r}, \vec{r'},\vec{q}}$, where $\vec{r}$ and  $\vec{r'}$ belong to the primitive cell of the crystal and $\vec{q}$ resides in the first Brillouin zone $\BZ$ of the crystal.
This Fourier transformation of the susceptibility is obtained from the convolution of two KKR Green's function over $\BZ$.
The spatial dependence of the susceptibility in the primitive cell is developed in the atomic centered coordinates (density indices $ij$ are suppressed for clarity) as
\begin{align}
\suscSymb\fbr{\vec{r}, \vec{r'}} = \sum_{LL'} 
    \suscSymb^{ss'}_{LL'}
    \fbr{r_{s},r_{s'}}
    \Ylm_{L }\fbr{\hat{\vec{r}}_{s }}
    \Ylm_{L'}\fbr{\hat{\vec{r}}_{s'}}
\end{align}
Here, $\vec{r}_{s}$ should be understood as a reduced vector $\vec{r}$, $\vec{r}_{s} \equiv \vec{r} - \vec{s}_{s}$, where $\vec{s}_{s}$ is the center of the atomic polyhedron containing $\vec{r}$.
The angular dependence of the spatial arguments in each cell is developed using spherical harmonics $\Ylm_{L}$ with index $L \equiv lm$ and argument $\hat{\vec{r}}_{s} \equiv \vec{r}_{s}/\abs{\vec{r}_{s}}$.
Note that in the case of full potential calculations, the product of two shape functions for the polyhedra $s$ and $s'$ is incorporated into the definition of $\suscSymb^{ss'}_{LL'}\fbr{r_{s},r_{s'}}$.
The radial dependence of $\suscSymb\fbr{r_{s},r_{s'}}$ can be developed in different ways.
In our previous works, cf., e.g., \cite{Buczek2011a}, we took advantage of Chebyshev polynomials in order to achieve a rapid convergence with respect to the number of basis functions, as,
typically, only around 12 were sufficient to fairly describe the dependence.
In the current work, the function $\suscSymb\fbr{r_{s},r_{s'}}$ is given directly in the real space on a Gaussian mesh between $0$ and the radius of the bounding atomic sphere $s$.
This approach leads to somewhat increased memory consumption but greatly enhances the fulfillment of the Goldstone sum rule.
Upon this step, the susceptibilities are represented as ordinary two-dimensional matrices $\suscSymb_{\lambda\lambda'}$ where the super-index $\lambda = \fbr{i,s,L,\rho}$ comprises the density channel designation $i$, site $s$, angular quantum numbers $L \equiv lm$, and the radial index $\rho$.
In this representation, Eq.~\eqref{eq:susceptibilityDysonEquation} can be solved by the following direct matrix inversion \cite{Staunton1999,Staunton2000,Thakor2003}:
\begin{align}
\suscSymb = \fbr{\onemat - \suscSymb\KS \fbr{K\xc + \vC}}^{-1} \suscSymb\KS.
\end{align}

One particular aspect of density response evaluation in NC systems turns out to become particularly algorithmically challenging compared to the collinear case, namely the evaluation of the trace over spin indices in Eq.~\eqref{eq:Sij}.
We recall that the non-local part of the KKR Green's function consists of a product of three spin-matrices such that there emerges a contribution to $S_{ij}$ as follows
\begin{align}
  c_{ij} = \tr 
  \sigma^{i}
    \nu_{0} \nu_{1} \nu_{2}
    \sigma^{j}
    \nu_{3} \nu_{4} \nu_{5}.
\label{eq:cij}
\end{align}
where the blocks $\nu_{m}^{\zeta}$ of each spin matrix
\begin{align}
\nu_{m} =     
    \begin{pmatrix}
        \nu_{m}^0 & \nu_{m}^2 \\
        \nu_{m}^3 & \nu_{m}^1
    \end{pmatrix}
\end{align}
are themselves matrices with the site, angular, and radial indices and $\tr$ stands for spin index trace.
To be specific, when the non-local contribution to the susceptibility is concerned, $\nu_{1}$ and $\nu_{4}$ represent the backscattering operators and the remaining $\nu_{m}$ denote regular single site scattering solutions. While there is conceptually no difficulty in performing the matrix multiplication and the trace operation directly, this brute force approach to the calculation of the spin matrix elements features two major disadvantages which ultimately would lead to serious deterioration of the numerical performance.
First, the blocks $\nu_{m}^\zeta$ can become sparse, e.g., concerning the angular indices when no full potential or spin-orbit corrections are taken into account.
The direct matrix multiplication would then produce terms known to be zero from the onset.
Second, a particular product of these blocks (out of 256 appearing in the trace)
\begin{align}
    \prod_{m = 0}^{5} \nu_{m}^{\zeta_{m}}
\label{eq:NuProd}
\end{align}
can contribute to several out of 16 different pairs of density indices $ij$. If Eq.~\eqref{eq:cij} were to be repeatedly evaluated for varying $i$ and $j$, this product defined by specific $\cbr{\zeta_{0},\ldots,\zeta_{5}}$ would be unnecessarily repeatedly evaluated multiple times as well.
In order to manage this complexity, we developed a symbolic algebra scheme for computing the contribution of different products to the given $ij$ pair.
Since the product in Eq.~\eqref{eq:NuProd} involves matrices, it is necessary to keep track of the order of $\nu_{m}^{\zeta_{m}}$ in it.
To this end, we devised a new method for handling noncommutative algebraic operations and implemented it using symbolic algebra language primitives of Mathematica \cite{WolframResearch2025}.
Subsequently, the symbolic results are used to automatically generate the FORTRAN code deployed in the high-performance computing of the traces.
A more detailed exposition of the algorithm is given in Appendix \ref{app:SymbolicHandling}.
This strategy feature several decisive advantages over the traditional manual implementation of scientific software.
It combines the ease of symbolic manipulation with the numerical performance of low-level programming languages without the necessity of the tedious manual implementation in the latter.
Furthermore, a thorough automatic validation of symbolic models is decisively easier than the one of the low level software.

\subsection{Antiferromagnetic magnon peak fitting}
\label{subsec:AntiferromagneticMagnonPeakFitting}

\begin{figure}
    \centering
	\includegraphics[width=0.48\columnwidth]{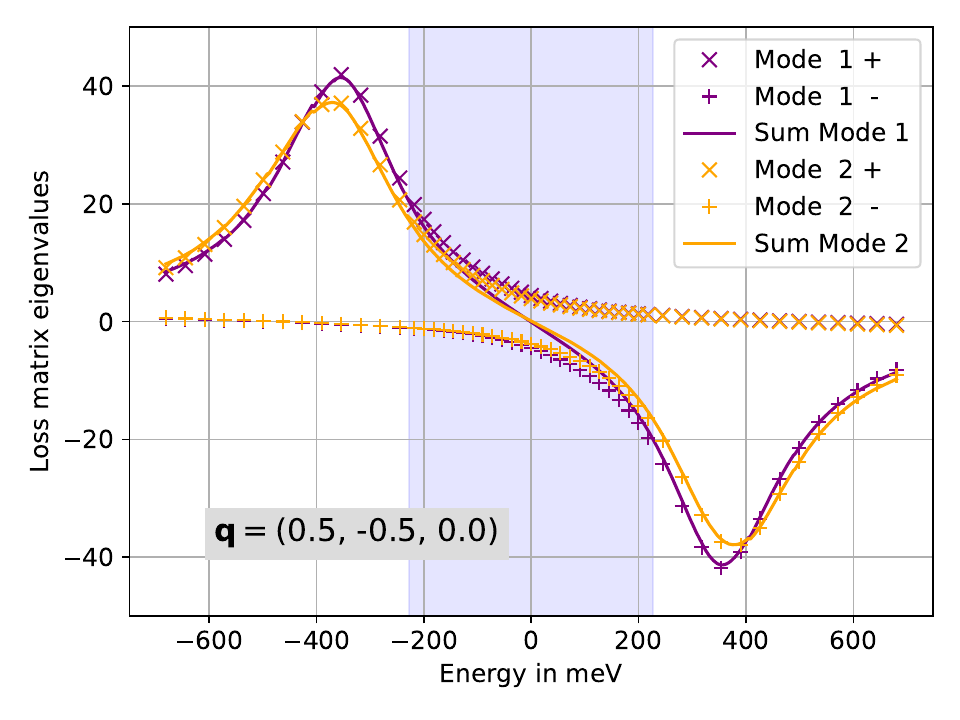}
	\caption{Magnon peak fitting in the case of enhanced susceptibility. $\times$, $+$, and $\cdot$ points correspond to the three relevant, i.e., of significant value, magnon mode eigenvalues of the loss matrix evaluated in our LRTDDFT scheme for energies on the nearly real axis. Following the properties of the Fourier-transformed response function, cf. Appendix ~\ref{app:KaellenLehmann}, the magnon peaks show typical anti-symmetry with respect to $\omega = 0$. In the blue region, the energy mesh density is increased to stabilize the peak parameters' fit. Note that the mesh density shown in the plot, for better visibility, is only a fraction of the actual density.}
\label{fig:DanishFit}
\end{figure}

As pointed out by Skovhus and Olsen \cite{Skovhus2022}, the magnon peak fitting in the case of the antiferromagnets is more challenging than for the ferromagnets, especially for small momenta.
It is so due to the characteristic presence of the two peaks anti-symmetric with respect to $\omega = 0$ becoming very close to each other in the short-wavelength regime, see Fig.~\ref{fig:DanishFit}.
Principally, we analyze and fit the magnon peaks on the nearly real axis, i.e., along a line on the upper complex semi-plane parallel to the real axis and in the distance $\gamma$ from it.
The susceptibility is directly evaluated for points $\omega + \ii \gamma$ and
no intermediate analytic continuation to real real axis ($\omega + \ii \zp$) is performed.
This is for two reasons.
First, the analytic continuation is known to be numerically ill-conditioned and unstable.
Despite extended research, we identified no reliable, robust, and automatable numerical algorithm for its execution.
Second, weakly damped magnons appear as poles very close to the real axis with corresponding significant eigenvalue magnitude and extremely narrow Lorenz-like peak practically precluding correct fit. 
In this work, referring to \cite{Skovhus2022}, we deploy the following magnon line ansatz
\begin{align}
    a_{\lambda}\fbr{\vec{q}, \omega} = 
    - A_{\lambda\vec{q}}
    \fbr{
        \frac{1}{
            \fbr{\omega - \omega_{\lambda\vec{q}}}^2 + 
            \fbr{\gamma + \eta_{\lambda\vec{q}}}^2}
        -
        \frac{1}{
            \fbr{\omega + \omega_{\lambda\vec{q}}}^2 + 
            \fbr{\gamma + \eta_{\lambda\vec{q}}}^2}
    }.
\label{eq:magnonPeakModel}
\end{align}
Here, $A_{\lambda\vec{q}}$ stands for the spectral weight of the magnon mode, $\omega_{\lambda\vec{q}}$ is the magnon frequency, and $\eta_{\lambda\vec{q}}$ the intrinsic magnon decay parameter. We recall $\gamma$ to be the distance of the nearly real axis, i.e., the artificial broadening. For large $\omega_{\lambda\vec{q}}$, the peak's maximum appears approximately at this frequency, and the full width at half-maximum (FWHM) of the peak amounts to roughly $2\eta_{\lambda\vec{q}}$. As carefully discussed in \cite{Skovhus2022}, for $\omega_{\lambda\vec{q}} \to 0$, the peak's maximum and FWHM do not directly yield these fit parameters.

On the nearly real axis, the eigenvalues of the $L(\omega + \ii\gamma)$ do not necessarily vanish for $\omega = 0$, contrary to the requirement of Eq.~\eqref{eq:magnonPeakModel}.
This difficulty can be circumvented upon considering the sum of eigenvalues corresponding to each other for positive and negative frequencies.
As evident from Fig.~\ref{fig:DanishFit}, the sum goes through the origin and is represented well by the equation.

Although the ansatz in Eq.~\eqref{eq:magnonPeakModel} includes only three parameters, the quality of the fit and the stability of the results for varying $\vec{q}$, especially in the long wavelength regime, improve considerably with the number of points in the fitting set.
The low frequency region (shaded blue region in Fig.~\ref{fig:DanishFit}) turns out to be particularly relevant and in our calculations we typically increase the frequency point density there.

\subsection{Convergence considerations}
\label{subsec:ConvergenceConsiderations}

\begin{figure}
	\centering
	\includegraphics[width=0.98\columnwidth]{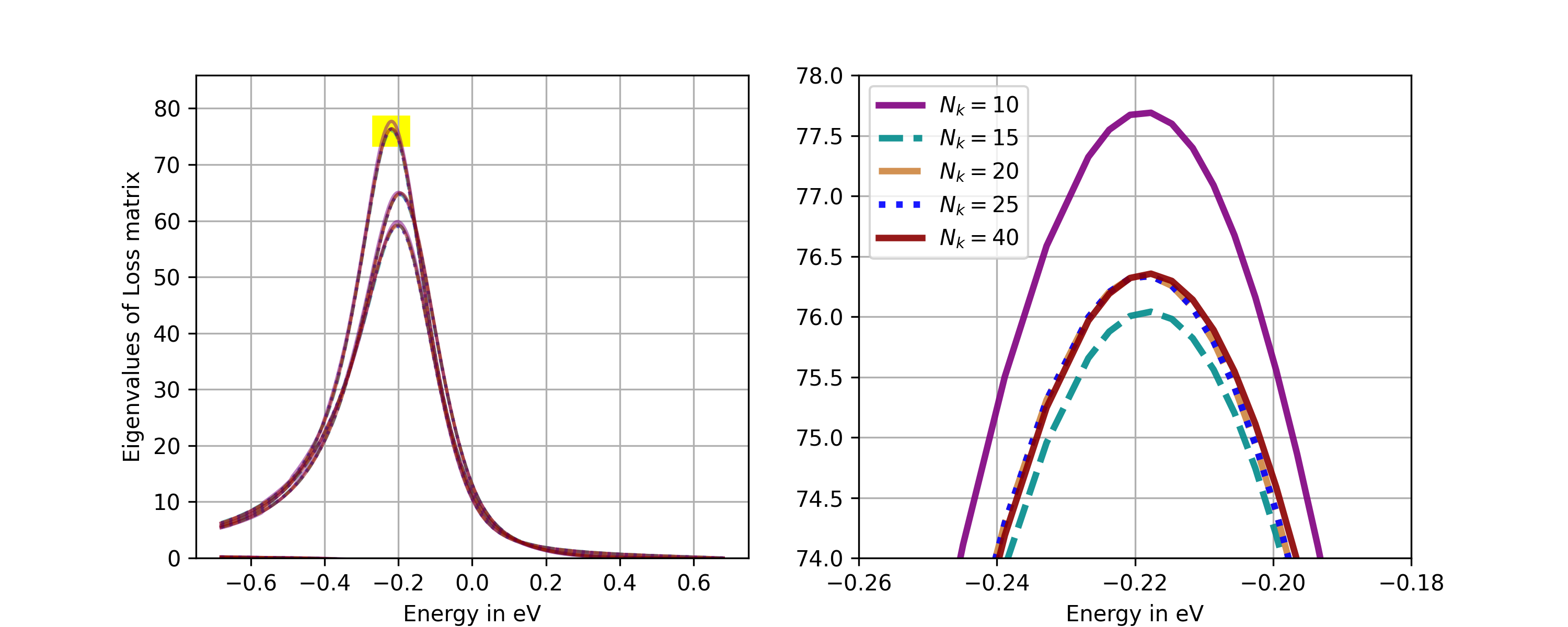}
	\caption{Convergence of the enhanced susceptibility with respect to the number of $\vec{k}$ points used for the Brillouin zone integration. For the case of Mn$_3$Ir, the most dense mesh necessary has $N_{k} \times N_{k} \times N_{k}$ points with $N_{k} = 25$. The three groups of lines correspond to the three relevant magnon eigenvalues.}
	\label{fig:kintegeration}
\end{figure}

Let us conclude this section with several remarks regarding the numerical convergence properties of our scheme.
They resemble much the ones encountered in the collinear case \cite{Buczek2011a}.
The most computationally expensive part of the algorithm is the Brillouin zone integration of the product of two backscattering operators, i.e., their $\vec{k}$-convolution.
For the case of Mn$_3$Ir, the most dense mesh necessary has $N_{k} \times N_{k} \times N_{k}$ points with $N_{k} = 25$, cf.~\ref{fig:kintegeration}.
Note that this number is reduced considerably once the backscattering operators are evaluated away from the real axis.
Furthermore, the number $N_{k} = 25$ is moderate compared to other systems.
For example, the bcc Fe requires $N_{k} = 100$ or more.
This is due to the fact, plainly speaking, that Fe features large density of states at the Fermi energy.
The strongly dispersing bands of Mn$_3$Ir crossing the Fermi level result in a flat and low density of states in this region and the relaxed convergence demand.

In the spherical harmonics expansion of the KKR Green's function it is enough to include only states up to $d$ ($l_\mathrm{max}^\mathrm{KKR} = 2$) when computing the density response functions.
When concentrating on the dynamics of moments as a whole, the expansion of the response functions with $l_\mathrm{max}^\mathrm{susc} = 0$ yields already accurate enough results.
The cases of $l_\mathrm{max}^\mathrm{susc} = 1$ and $l_\mathrm{max}^\mathrm{susc} = 2$ are indiscernible and differ only weakly from the spherical approximation.

We find, however, that the radial dependence of susceptibility components in atomic cell influences the accuracy of the calculation, in particular the Goldstone divergence, strongly.
The radial mesh points are chosen according to the Gauss-Legendre quadrature scheme. 
A Neville interpolator is used in order to retrieve the values of the radial solutions on the Gauss mesh from the denser logarithmic mesh which is used in the underlying KKR calculations.
It turns out that around 20 points per atom balance accuracy and computational cost reasonably in the case of Mn$_3$Ir.

\section{Results}
\label{sec:Results}

\subsection{Electronic ground state}
\label{subsec:ElectronicGroundState}

\begin{figure}
	\centering
	\includegraphics[width=0.45\textwidth,trim={25 150 25 150},clip]{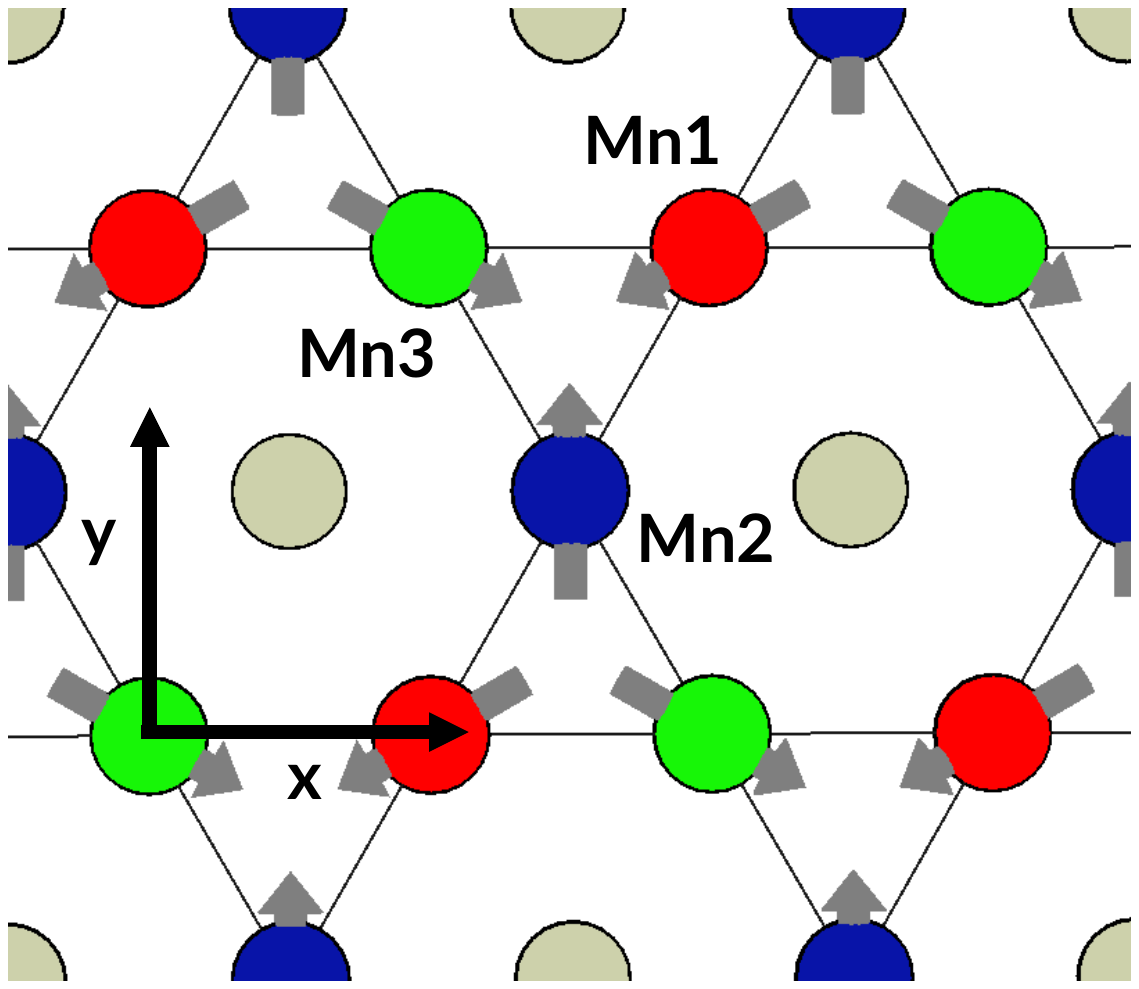}
	\caption{
$111$-plane of Mn$_{3}$Ir revealing the characteristic non-collinear KAFM magnetic pattern.
The staking of these shifted planes generates the entire crystal.
The color coding of the magnetic Mn atoms is consistently maintained throughout the paper.
The global coordinate system is oriented such that the $z$-axis points out of the $111$-plane and $x$ and $y$ axes lie in this plane.
The local coordinate systems for each Mn atom have their $z$-axes along the moments' directions,
the $y$-axes are assumed to lie in the kagome $111$-plane,
and $x$-axes are chosen such that right-handed coordinate systems form.
The Mn2 atomic magnetic moment is taken to be oriented in the global $y$ direction.
}
	\label{fig:structure}
\end{figure}

Fig.~\ref{fig:structure} depicts the kagome $111$-plane of Mn$_{3}$Ir investigated in this paper.
The lattice constant of the underlying cubit lattice is taken to be $a = 3.785$\AA.
The Mn-Mn atomic distance amounts to $a/\sqrt{2}$
and the distance between 111-planes to $a/\sqrt{3}$.
The global coordinate system is oriented such that $z$-axis points out of the $111$-plane and $x$ and $y$ axes lie in this plane.
The local coordinate systems for each Mn atom have their $z$-axes along the moments' directions,
the $y$-axes are assumed to lie in the kagome $111$-plane,
and $x$-axes are chosen such that right-handed coordinate systems form.
The Mn2 atomic magnetic moment is taken to be oriented in the global $y$ direction.

\begin{figure}
	 \centering
	\includegraphics[width=0.7\textwidth,trim={0 0 0 0},clip]{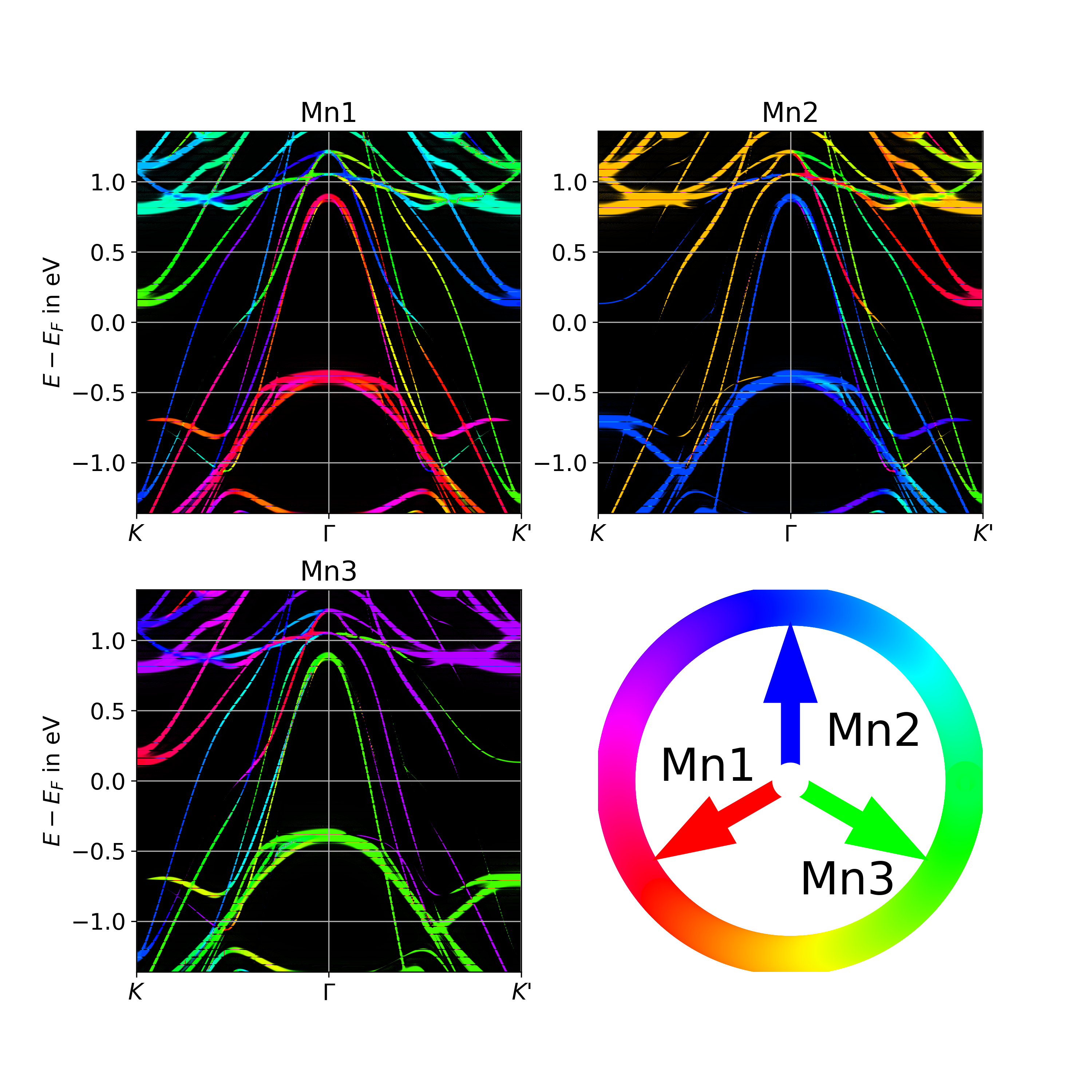}
	\caption{Electronic Bloch spectral function showing the contributions of the Mn atoms to the total band structure. The directions within the Brillouin zone can be found as an inset in Fig.~\ref{fig:dispersionIrMn3}. Color encodes information about the spin polarization (magnetization direction) of the Bloch state for a particular electronic momentum $\vec{k}$. The intensity of the band signifies the Bloch amplitude of the state on the atom.}
	\label{fig:bsf_color}
\end{figure}

The resulting band structure is intrinsically spin-polarized despite its vanishing net magnetization
\cite{%
gurungNearlyPerfectSpin2024b,%
huSpinHallEdelstein2025%
},
cf.\ Fig.~\ref{fig:bsf_color},
and the altermagnetic symmetry of the band structure is clearly discernible.
It originates from the fact that magnetic moments of different directions are embedded in different crystal environments imposed by the kagome pattern.
The altermagnetic symmetry has one more consequence:
The amplitude of the Bloch states (electronic occupation and spin polarization) on particular magnetic atoms can differ even for wave vectors related by the symmetry of the underlying non-magnetic lattice.
This is clearly seen in Fig.~\ref{fig:bsf_color} for Mn2 and Mn3 atoms as asymmetry between $K$ and $K'$ directions.
At $\Gamma$, one clearly identifies occupied electronic bands around 400meV below the Fermi level polarized along the respective moments' directions.
They are responsible for the moments' formation.
The Fermi surface is formed by strongly dispersing bands of heterogeneous spin polarizations.
Above the Fermi level and away from the center of the Brillouin zone, the magnetic orientation of bands differs strongly from the moment direction.
In our recent study \cite{Eilmsteiner2026a}, we have shown that this peculiarities of the NC altermagnetic band structure result in an intricate mode-selective Landau magnon damping.

\subsection{Spectral analysis of density excitations}
\label{subsec:SpectralAnalysis}

\begin{figure}
	\centering
	\includegraphics[width=1\columnwidth]{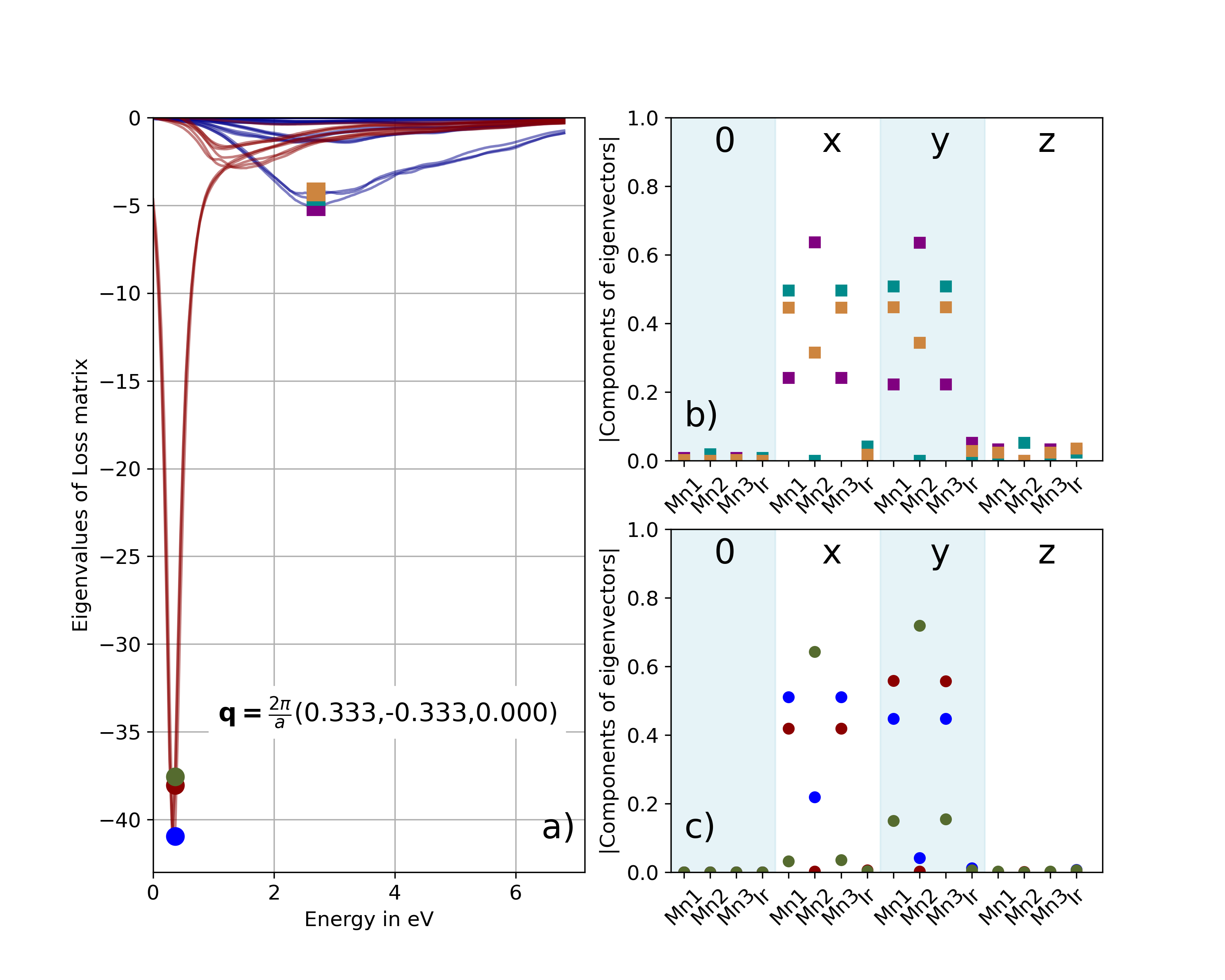}
	\caption{Eigenvalues (a) and some prominent eigenvectors (b,c) of the Loss matrix for a wave vector $\mathbf{q}$ in the middle of the Brillouin zone. The eigenvalues of the KS system shown in blue form a more flat spectrum while the eigenvalues of the interacting system, in red, exhibit a strong magnon peak. Analysis of the eigenvectors, given in the site local basis, shows the Heisenberg-like dynamical behavior of the interacting system, while the prominent eigenvectors of the KS system have non vanishing charge (0) and longitudinal (z) components.}
	\label{fig:KSSpectrumExample}
\end{figure}

The spectrum of particle-hole and collective excitations is given by the loss matrix of, respectively, KS and enhanced susceptibility, cf.\ Eq.\ \eqref{eq:lossMatrix}.
Owing to the clear interpretation of the eigenvalues $a_{\lambda}\fbr{\vec{q}, \omega}$ of
$L^{ij}\fbr{\vec{r}, \vec{r}', \vec{q}, \omega}$ we take them,
in most cases, directly as the starting point for any subsequent analysis.
Fig.~\ref{fig:KSSpectrumExample} presents an example of the KS particle-hole spectrum for a selected momentum $\vec{q}$.
The series of data points for a given $\omega$ represents the eigenvalues $a_{\lambda}\fbr{\vec{q}, \omega}$ indexed by $\lambda$, each associated with a different eigenvector $\xi_{\lambda}^{i}\fbr{\vec{r}, \vec{q}, \omega}$.
For given $\fbr{\vec{q}, \omega}$-pair, the set of $\cbr{\xi_{\lambda}^{i}\fbr{\vec{r}}}$ forms an orthogonal and complete basis for representing charge and spin densities in the primitive cell of the crystal.
In general, when arguments $\fbr{\vec{q}, \omega}$ change, a new set of eigenvectors is obtained.
The index $\lambda$ runs between 1 and $\chi$-matrix order given by the choice of computational basis, roughly of order of $10^{3}$ for our kagome magnet.
In Fig.~\ref{fig:KSSpectrumExample}, all eigenvalues with the magnitude over certain threshold are shown, the remaining being of minor weight.
Still, we observe that the KS excitations, depicted in blue in figure \ref{fig:KSSpectrumExample}a), are associated with a multitude of relevant eigenvectors (figure \ref{fig:KSSpectrumExample} b)) involving, in general, charge and magnetization components of different directions.
The picture changes when the true (enhanced) response is considered, cf. red lines in ~Fig.~\ref{fig:KSSpectrumExample}a), markers in ~Fig.~\ref{fig:KSSpectrumExample}c), and the data in Fig.~\ref{fig:DanishFit}.
In the energy window below 500meV, only a few eigenvalues of the enhanced susceptibility show a sizable magnitude, other being negligible.
Fig.~\ref{fig:KSSpectrumExample}c) reveals that the corresponding modes practically  involve only oscillating spin density oriented perpendicular to the local ground state magnetization direction, whereas the longitudinal and charge components exhibit vanishing weight. This is in good agreement with a basic assumption of the atomistic Heisenberg model, in which atomic magnetic moments are assumed to be of constant magnitude and charge degrees of freedom are neglected.   


In this vein, the magnons known from the Heisenberg model arise naturally in the dynamical susceptibility calculations.
The three relevant eigenvectors correspond to the three magnon modes and will be discussed more carefully below.
Interestingly, the eigenvalues for different frequencies $\omega$ can relatively easily be assigned to a particular mode
because the magnon eigenvector $\xi_{\lambda}^{i}\fbr{\vec{r}, \vec{q}, \omega}$ changes only weakly with the frequency.
(However, we recall that for a given frequency $\omega$ the magnon eigenvectors $\xi_{\lambda}$ are orthogonal and thus much different.)
We finally note that for energies above the magnon range, there additionally emerges a plethora of coupled charge and magnetization density excitations, e.g., longitudinal spin fluctuations and plasmons \cite{Buczek2020}.
They will be studied in a separate work.

\subsection{Magnon dispersion and the chirality-selective Landau damping}
\label{subsec:magnonDispersion}

\begin{figure}
	\includegraphics[width=0.95\textwidth]{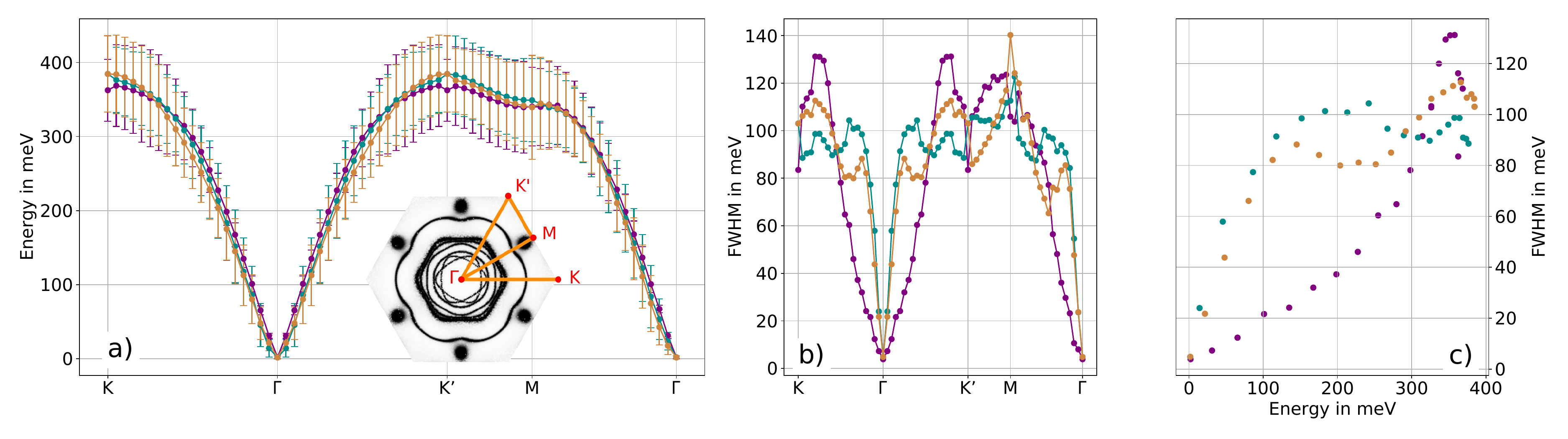}
	\caption{
The spectrum of magnons in the kagome 111-plane of Mn$_3$Ir. The color-coding of the magnon branches is used consistently in this section.
(a) Magnon energies as functions of momentum. The tree branches are practically degenerate concerning the energies but not the mode-selective damping.
Error bars signify the FWHM of the magnon peak, i.e. the inverse lifetime of the quasiparticle.
In the inset, the 111-cut of the Fermi surface is shown together with the path through the Brillouin zone.
(b) FWHM of the magnon peaks as a function of momentum.
(c) FWHM of the magnon peaks as a function of the magnon energy.
The mode-selective Landau attenuation is evident from Figs. b) and c).
}
\label{fig:dispersionIrMn3}
\end{figure}

Figure \ref{fig:dispersionIrMn3} shows the magnon dispersion and the inverse lifetimes (shown as bars designating the full with at half-maximum, FWHM, of the magnon peak) obtained from our LRTDDFT calculations. 
Three linearly dispersing magnon modes emerge in the long wave-length limit from three distinct Goldstone modes, as expected from the symmetry analysis outlined above.
Their spin wave velocities amount to around $v_\Gamma\approx 80$km/s
and their maximum energies at the edge of the Brillouin zone reach about 370meV.
The modes are clearly Landau damped but remain well-defined for all momenta
and can be treated as underdamped harmonic oscillators.
The maximal FWHM of a magnon mode is found to be about 140meV but most magnons live significantly longer.
While the magnon dispersions of Mn$_3$Ir and Mn$_3$Rh are qualitatively similar, the magnons in the first system are of higher energies and stronger Landau damped than in the latter, cf.~\cite{Eilmsteiner2026a}.

Similarly to the case of Mn$_3$Rh investigated recently \cite{Eilmsteiner2026a}, above the threshold of about 150meV, the FWHM of magnons peaks in Mn$_3$Ir is a non-monotonous function of both momentum and magnon energy.
This makes these two materials extremely promising candidates for the terahertz spintronic applications.
The relatively low damped high-frequency modes, before their decay, perform a larger number of revolutions compared to their low-frequency companions,
despite featuring larger attenuation parameter (FWHM).
This feature will be further investigated in Sec.~\ref{subsec:magneticModes}.

Another remarkable property of the Landau damping shared by the two kagome magnets is its pronounced mode selectivity.
The decay rates differ significantly even for magnonic modes of practically the same energy and momentum.
To be specific, let us consider the low-momentum magnons in Fig.~\ref{fig:dispersionIrMn3}b)
Two modes ($\alpha$-type, green and yellow markers) are up to factor 3 weaker Landau damped than the third branch of $\beta$-type (magenta).
As discussed carefully before \cite{Eilmsteiner2026a}, this selectivity can be traced to the spectrum of the non-collinear Stoner excitations resulting from the altermagnetic character of the underlying electronic band structure interplaying with the forms of the magnons' eigenvectors.

\subsection{Magnetic modes}
\label{subsec:magneticModes}

\begin{figure}
	\centering
	\includegraphics[width=0.85\columnwidth]{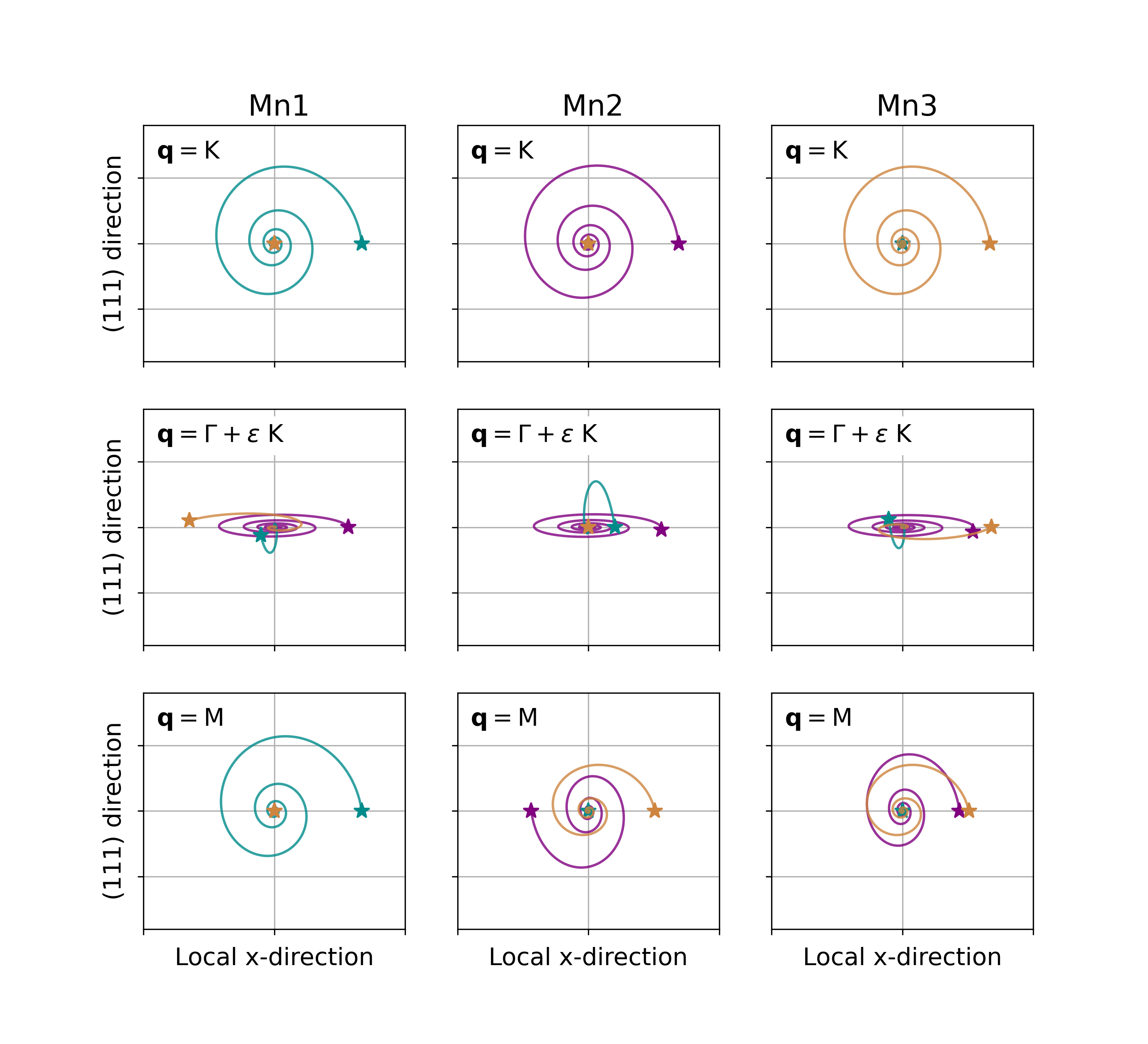}
	\caption{
Real-space and parametric real-time depiction of the magnetic moments' precession associated with magnon eigenvectors obtained from our LRTDDFT calculations, including the Landau attenuation.
The modes become excited at $t = 0$.
The three magnetic Mn atoms of the primitive cell are considered.
The color coding corresponds to the magnon branches identified in the dispersion given in Fig.~\ref{fig:dispersionIrMn3}
Stars signify the starting orientation at $t = 0$.
For $t > 0$, the components of the moments' deviations in the respective local moment coordinate systems are plotted.
}
	\label{fig:realspacemodes}
\end{figure}

\begin{figure}
	\centering
	\includegraphics[width=0.85\columnwidth]{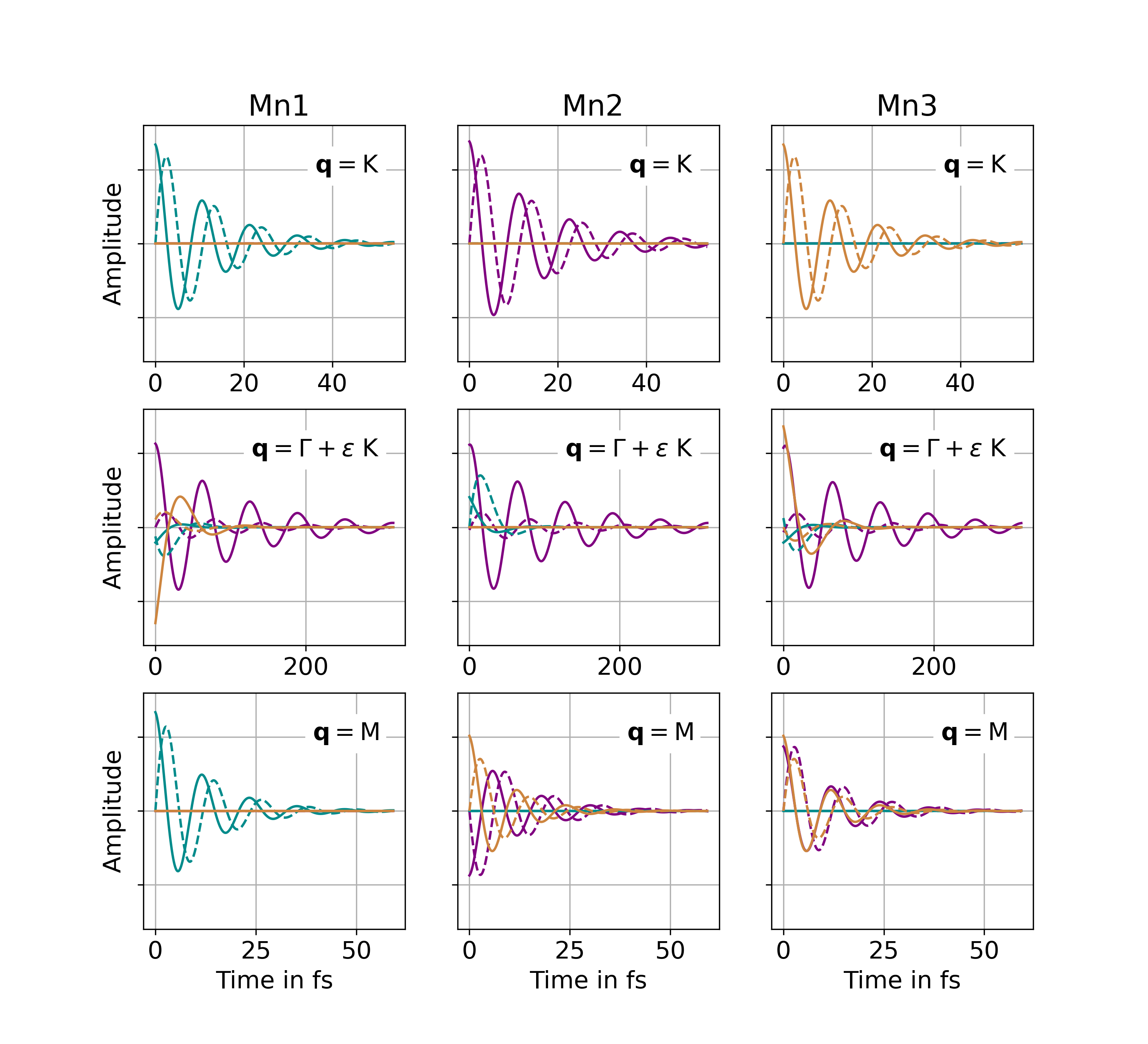}
	\caption{
The same patterns of precession as in Fig.~\ref{fig:realspacemodes} but shown as explicit function of time.
The solid lines stand for the local $x$ direction (out of 111-plane) while the dashed ones for the local $y$ direction (in 111-plane).  
}
\label{fig:realspacemodesAmplitude}
\end{figure}

Let us discuss the real-space and real-time dynamics of magnons modes identified in the preceding subsection.
In our analysis, we choose three representative momenta.
$K$ and $M$ lie at the edge of the Brillouin zone.
$K$ is equivalent to $K'$ by the altermagnetic symmetry.
The momentum $\Gamma + \epsilon K$ is located close to the zone center slightly off in the direction of $K$.

Let us consider first the Fig.~\ref{fig:realspacemodes}.
The lines stand for the time-parametrized Landau-damped precession of natural magnon modes at given momentum excited at $t = 0$.
For a particular atomic magnetic moment, the observer is positioned against the direction of the moment in the ground state.
The stars denote the initial direction of the moments at $t = 0$.
All three magnon modes are shown and kept apart using the color-coding from Fig.~\ref{fig:dispersionIrMn3}.
Fig.~\ref{fig:realspacemodesAmplitude} has the same physical content but the local $x$ and $y$ magnon components are shown as an explicit function of time.

Several features of these precession patterns point to the strong potential of KAFM as terahertz spintronic materials.
The small momentum ($\Gamma + \epsilon K$) eigenvectors involve the precession of all magnetic moments.
At the same time, they illustrate the pronounced chirality-dependent Landau damping in this class of systems.
We clearly observe that the $\beta$-mode performs numerous revolutions prior to its decay while the $\alpha$-modes complete hardly a single one before fading away.
This behavior is consistent with the FHWM discussion presented in Subsec.~\ref{subsec:magnonDispersion}.
Another interesting property of the magnonic eigenvectors is the magnetic sublattice localization for large momenta.
At the $M$-point, one of the modes localizes on Mn1 atoms while the other two excite Mn2 and Mn3 sublattices.
The phenomenon is even more pronounced at the $K$ point where in every mode only one sublattice involves in the precession.
This behavior can be used to gain precise control over the dynamics of single atomic moments.

\section{Summary}
\label{sec:Summary}

We presented a novel implementation of the \textit{ab initio} linear response time-dependent density functional theory addressing spin excitations in non-collinear magnets based on the Korringa-Kohn-Rostoker Green's function method.
We exposed the details of the numerical implementation, focusing on the managing the algorithmic complexity by means of symbolic computational algebra, and
discussed the generalization of the adiabatic local spin density approximation to the exchange-correlation kernel to the non-collinear case.
The emergence of multiple Nambu-Goldstone bosons in the non-collinear dynamic susceptibility calculations is elaborated on formally and from the numerical convergence point of view,
being linked to the zero-energy magnetic deformation eigenvectors.
With the aid of the formalism, we analyzed the spin density response of the KAFM IrMn$_3$,
in particular, the dispersion and Landau damping of magnons as well as the real-time and real-space dynamics of the spin-wave modes.
Due to the non-monotonous dependence of the damping on the magnon frequency and chirality
the excitations become attractive in the terahertz spintronics applications.
We hope that this work will give rise to further theoretical and experimental progress in the young field of magnonics based on non-collinear ground state systems.

\vspace{1cm}

\textit{Acknowledgments}. We thank Leonid M. Sandratskii, L\'a{}szl\'o{} Szunyogh, Patrick Perndorfer, Sebastian Paischer, and Igor Maznichenko for fruitful discussions.
P.B. and A.E. acknowledge the funding, respectively, by \"Osterreichischer
Fonds zur F\"orderung der Wissenschaftlichen Forschung (FWF)
under grant I 5384 and by Deutsche Forschungsgemeinschaft (DFG) under grant BU 4062/1-1.

\newpage
\appendix

\section%
[K\"all\'en-Lehmann spectral representation and the fluctuation-dissipation theorem]%
{K\"all\'en-Lehmann spectral representation and the \\fluctuation-dissipation theorem}
\label{app:KaellenLehmann}

The numerical evaluation of the density response functions is a notoriously difficult undertaking and the following relations
\cite{%
Nyquist1928,
Callen1951,%
Fetter1971,%
Kubo1966,%
Abrikosov1975,%
Giuliani2005}
appeared useful in the validation of our computer implementation and the interpretation of results.
The relations given below are completely general and apply even to the cases of non-collinearity non-centrosymmetric crystals with relativistic effects taken into account.

In order to simply the notation, we consider the response function between two operators labelled by $\alpha$. To be specific, in our case
\begin{align}
    \alpha = \fbr{i\vec{x}}
\end{align}
but the following consideration apply to any Hermitian operators being time-independent in the Schr\"odinger picture.
In order to study the response functions in periodic solids, we will nevertheless assume that the operators are ascribed to a particular primitive cell $\vec{R}$.
The retarded response function $\chi\ret$ is now defined as
\begin{align}
\chi_{\alpha\beta}\ret\fbr{\vec{R} - \vec{R}', t - t'} \equiv 
    - \ii \Heav\fbr{t - t'}
      \av{
          \comm{\hat{o}_{\alpha\vec{R}}\fbr{t}}{\hat{o}_{\beta\vec{R}'}\fbr{t'}}{}
      }. 
\end{align}
Since the function connects the change of the observable physical (real) observable represented by $\hat{o}_{\alpha\vec{R}}$ due to the disturbing physical field $\hat{o}_{\beta\vec{R}'}$ and as such it is a real scalar, $\chi_{\alpha\beta}\ret\fbr{\vec{R}, t} \in \Reals$. Consequently, the time and lattice Fourier transformation (cf.~Eq.~\eqref{eq:LatticeFourierTransformation}) obeys the \textit{conjugation symmetry}
\begin{align}
    \chi_{\alpha\beta}\ret\fbr{- \vec{q}, - \omega}^{\ast} = 
    \chi_{\alpha\beta}\ret\fbr{  \vec{q},   \omega}.
\end{align}
We remark that the above formula applies also to the case when the response function is evaluated not directly above the real axis ($\gamma > 0$ in Eq.~\eqref{eq:Schmalian})

The Fourier-transformed susceptibility is conveniently expressed in the \textit{K\"all\'en-Lehmann spectral representation} as
\begin{align}
\chi_{\alpha\beta}\ret\fbr{\omega, \vec{q}} =
\sum_{\vec{R} \in \latt} 
    \ee{ - \ii \vec{q} \cdot \vec{R}}
    \sum_{sf} \frac{\ee{ - E_{s}/(\kB T)}}{Z}
    \fbr{
        \frac{\braket{s|\hat{o}_{\alpha\vec{R}}|f}
              \braket{f|\hat{o}_{\beta \vec{0}}|s}}
             {\omega - \fbr{E_{f} - E_{s}} + \ii\zp}
        -
        \frac{\braket{s|\hat{o}_{\beta \vec{0}}|f}
              \braket{f|\hat{o}_{\alpha\vec{R}}|s}}
             {\omega + \fbr{E_{f} - E_{s}} + \ii\zp}
    },
\end{align}
where $\ket{s}$ is the \textit{many-body} electronic eigenstate of the unperturbed Hamiltonian with eigenenergy $E_{s}$ and $Z$ stands for the partition function
($Z\equiv\ee{-\Omega/(\kB T)}$, where $\kB$ is the Boltzmann constant and $T$ denotes the temperature).

One quantifies the natural (spontaneous) fluctuations of the ground state observables using the following \textit{correlation function}
\begin{align}
\crrl_{\alpha\beta}\fbr{\vec{R}, t} \equiv 
    \av{\hat{o}_{\alpha\vec{R}}\fbr{t}\hat{o}_{\beta\vec{0}}}
\end{align}
Fourier-transformed as
\begin{align}
\crrl_{\alpha\beta}\fbr{\vec{q}, \omega} = 2 \pi
\sum_{\vec{R} \in \latt} 
    \ee{ - \ii \vec{q} \cdot \vec{R}}
\sum_{sf} \frac{\ee{ - E_{s}/(\kB T)}}{Z}
    \braket{s|\hat{o}_{\alpha\vec{R}}|f}
    \braket{f|\hat{o}_{\beta \vec{0}}|s}
    \delta\fbr{\omega - \fbr{E_{f} - E_{s}}}.
\end{align}
Using the loss matrix
\begin{align}
L_{\alpha\beta}\fbr{\vec{q}, \omega} \equiv \frac{1}{2\ii}
    \fbr{
        \chi\ret_{\alpha\beta }\fbr{\vec{q}, \omega} - 
        \chi\ret_{\beta \alpha}\fbr{\vec{q}, \omega}^{\ast}
    }
\end{align}
the correlation function is written as
\begin{align}
\crrl_{\alpha\beta}\fbr{\vec{q}, \omega} =
    - \frac{2}{1 - \ee{ - \omega/(\kB T)}}
    L_{\alpha\beta}\fbr{\vec{q}, \omega}.
\label{eq:corrFromL}
\end{align}
$\crrl$, similar to $L$, is a Hermitian matrix
\begin{align}
\crrl_{\alpha\beta }\fbr{\vec{q}, \omega} =
\crrl_{\beta \alpha}\fbr{\vec{q}, \omega}^{\ast}.
\end{align}

One also proves that
\begin{align}
L_{\alpha\beta}\fbr{\vec{q}, \omega = 0} = 0
\label{eq:w0Hermitian}
\end{align}
which in turn implies that $\chi_{\alpha\beta}\ret\fbr{\vec{q}, \omega = 0}$ is a Hermitian and $\chi_{\alpha\beta}\ret\fbr{\vec{q} = \vec{0}, \omega = 0}$ a symmetric real matrix.

Let us assume that our system is weakly and mono-chromatically perturbed by a Hamiltonian of the following form
\begin{align}
\hat{H}\fbr{t} = \sum_{\alpha} 
    f_{\alpha} 
    \hat{o}_{\alpha\vec{R}}
    \cos\fbr{\vec{q} \cdot \vec{R} - \omega t - \phi}.
\end{align}
It is given in the Schr\"odinger picture. $f_{\alpha}$ are real scalars representing the potentials coupling to operators $\hat{o}_{\alpha}$. The power absorbed from this perturbing field reads
\begin{align}
P = - \frac{\omega}{2} 
    \sum_{\vec{R}, \vec{R} \in \latt}
        \ee{ - \ii \vec{q} \cdot \fbr{\vec{R} - \vec{R}'}}
    \sum_{\alpha\beta} 
        f_{\alpha}
        L_{\alpha\vec{R}\beta\vec{R}'}\fbr{\omega}
        f_{\beta}
  = - N \frac{\omega}{2} \sum_{\alpha\beta}
      f_{\alpha} 
      L_{\alpha\beta}\fbr{\vec{q}, \omega}
      f_{\beta}.
\label{eq:dissipationFromL}
\end{align}
In the above equation, $N$ stands for the number of primitive cells of the crystal under perturbation. As expected, $P$ is an extensive quantity. We observe that the absorption is given by the eigenvalues and eigenvectors of $L$. Since the absorption occurs due to the generation of the system's excited states, the eigensystem of $L$ yields them directly.

The fluctuation-dissipation theorem reflects the observation that both the spontaneous fluctuations of the ground state observables (Eq.~\eqref{eq:corrFromL}) as well as the absorption of power from external driving field (Eq.~\eqref{eq:dissipationFromL}) are given by the anti-Hermitian part of the response function $L$.

Furthermore, $\crrl_{\alpha\beta}\fbr{\vec{q}, \omega}$ allows to write down the Kramers-Kronig relation for the response function \cite{Kronig1926} as
\begin{align}
  \chi_{\alpha\beta}\ret\fbr{\vec{q}, \omega} =&
     \int \frac{d\omega'}{2\pi} 
     \frac{\crrl_{\alpha\beta}\fbr{\vec{q}, \omega'}}
         {\omega - \omega' + \ii\zp}
    -\int \frac{d\omega'}{2\pi} 
     \frac{\crrl_{\beta\alpha}\fbr{ - \vec{q}, \omega'}}
         {\omega + \omega' + \ii\zp}.
\end{align}
The above allows to show that
\begin{align}
\crrl_{\alpha\beta }\fbr{   \vec{q},   \omega}
-
\crrl_{\beta \alpha}\fbr{ - \vec{q}, - \omega}
=
- 2 L_{\alpha\beta }\fbr{   \vec{q},   \omega}.
\end{align}

Finally, for completeness, we write down equivalent expressions for the advanced ($\chi\adv$), causal ($\chi$), and thermal ($\chi\tpr$) response functions
\begin{align}
\chi_{\alpha\beta}\adv\fbr{\vec{R} - \vec{R}', t - t'} &\equiv 
      \ii \Heav\fbr{t' - t}
      \av{
          \comm{\hat{o}_{\alpha\vec{R} }\fbr{t}}
               {\hat{o}_{\beta \vec{R}'}\fbr{t'}}{}
      }. \\
\chi_{\alpha\beta}    \fbr{\vec{R} - \vec{R}', t - t'} &\equiv 
    - \ii 
      \av{
          \tOrd{\hat{o}_{\alpha\vec{R} }\fbr{t}
                \hat{o}_{\beta \vec{R}'}\fbr{t'}}{}
      } \\
\chi_{\alpha\beta}\tpr
    \fbr{\vec{R} - \vec{R}', \tau - \tau'} &\equiv 
    - \av{
          \tOrd{\hat{o}_{\alpha\vec{R} }\fbr{\tau }
                \hat{o}_{\beta \vec{R}'}\fbr{\tau'}}{}
      }.
\end{align}
The corresponding spectral representations for the first two read
\begin{align}
  \chi_{\alpha\beta}\adv\fbr{\vec{q}, \omega} =&
      \int \frac{d\omega'}{2\pi} \frac{\crrl_{\alpha\beta}\fbr{\vec{q}, \omega'}}{\omega - \omega' - \ii\zp}
     -\int \frac{d\omega'}{2\pi} \frac{\crrl_{\beta\alpha}\fbr{ - \vec{q}, \omega'}}{\omega + \omega' - \ii\zp}, \\
  \chi_{\alpha\beta}    \fbr{\vec{q}, \omega} =&
      \int \frac{d\omega'}{2\pi} \frac{\crrl_{\alpha\beta}\fbr{\vec{q}, \omega'}}{\omega - \omega' + \ii\zp}
     -\int \frac{d\omega'}{2\pi} \frac{\crrl_{\beta\alpha}\fbr{ - \vec{q}, \omega'}}{\omega + \omega' - \ii\zp}.
\end{align}
On the other hand, the Matsubara representation of $\chi\tpr$ contains only bosonic frequencies $\omb{m} = 2 m \pi \kB T$, and its Fourier transformation and the corresponding spectral representation reads
\begin{align}
  \chi_{\alpha\beta}\tpr\fbr{\vec{q}, \omb{m}} &\equiv
    \int_{0}^{1/\fbr{\kB T}} \chi_{\alpha\beta}\tpr\fbr{\vec{q}, \tau} \ee{\ii\omb{m}\tau} \\
    &=
      \int \frac{d\omega}{2\pi} \frac{\crrl_{\alpha\beta}
      \fbr{   \vec{q}, \omega}}{\ii\omb{m} - \omega}
     -\int \frac{d\omega}{2\pi} \frac{\crrl_{\beta\alpha}
      \fbr{ - \vec{q}, \omega}}{\ii\omb{m} + \omega}.
\end{align}

\section{The symbolic handling of the spin matrix traces}
\label{app:SymbolicHandling}

The evaluation of the dynamical susceptibility based on the KKR Green's functions involves performing the spin index trace of the product of multiple matrices in the typical form of the one in Eq.~\eqref{eq:cij}.
To this end, we devised a method for handling noncommutative algebraic operations motivated by the G\"odel numbering developed by him in the context of the proof of his incompleteness theorem \cite{Goedel1931}. The specific implementation is formulated in the symbolic algebra language primitives in Mathematica \cite{WolframResearch2025}.
The matrices $\nu_{m}$ are represented symbolically as
\begin{align}
\nu_{m} =     
    \begin{pmatrix}
        x^{0 \cdot 4^m} & x^{2 \cdot 4^m} \\
        x^{3 \cdot 4^m} & x^{1 \cdot 4^m}
    \end{pmatrix},
\label{eq:num}
\end{align}
where $\cdot$ denotes ordinary multiplication.
The matrix multiplication of such matrices 
\begin{align}
    \prod_{m = 0, 1, \ldots} \nu_{m}
\end{align}
and subsequent trace taking generates polynomials of a form allowing for an automatic identification of the original position of a specific factor in the above product.
It relies on the fact that the power of $x$ expressed in quaternary numeral system can be directly interpreted as the sought-after order.

As a simpler example, we analyse the charge and spin density calculations from the KKR Green's function having the form of
\begin{align}
  \rho^{i} = \tr 
  \sigma^{i} \nu_{0} \nu_{1} \nu_{2}.
\label{eq:rhoi}
\end{align}
When the form given by Eq.~\eqref{eq:num} is inserted, one obtains the following column vector for the density channels
\begin{align}
\begin{pmatrix}
\rho^{0} \\
\rho^{x} \\
\rho^{y} \\
\rho^{z}
\end{pmatrix}
=
\begin{pmatrix}
x^{56}+x^{54}+x^{45}+x^{35}+x^{27}+x^{21}+x^{14}+1               \\
x^{59}+x^{53}+x^{46}+x^{32}+x^{24}+x^{22}+x^{13}+x^3             \\
- \ii \fbr{x^{59}+x^{53}-x^{46}-x^{32}-x^{24}-x^{22}+x^{13}+x^3} \\
x^{56}+x^{54}-x^{45}-x^{35}-x^{27}-x^{21}+x^{14}+1
\end{pmatrix}
\end{align}
To be specific, let us consider the $x^{24}$ term.
In the quaternary representation $24_{10} = 120_{4}$ and corresponds element $1$ of $\nu_{2}$ ($\nu_{2}^{22}$), element $2$ of $\nu_{1}$ ($\nu_{1}^{12}$), and element $0$ of $\nu_{0}$ ($\nu_{0}^{11}$).
We see that this product appears in the expression for $\rho^{x}$ and $\rho^{y}$ spin density with the coefficients of $1$ and $\ii$, respectively.
In other words, in this specific calculation, the product 
\begin{align}
    \nu_{0}^{11} \nu_{1}^{12} \nu_{2}^{22}
\end{align}
is computed only once but used twice, in the sum yielding $\rho^{x}$ (contribution $\nu_{0}^{11} \nu_{1}^{12} \nu_{2}^{22}$) and $\rho^{y}$ (contribution $\ii\nu_{0}^{11} \nu_{1}^{12} \nu_{2}^{22}$). 
The generalization to the case of products containing $6$ spin matrices, cf. Eq.~\eqref{eq:cij}, follows the same pattern.
In this case, one obtains $256$ different $6$-products of $\nu$'s, each used $4$ times in different $ij$-pairs of spin channels.
With the determined contributions and coefficients at hand, the automatic generation of the FORTRAN computer code is facilitated.
\printbibliography
\end{document}